\documentclass[12pt,a4paper]{article}
\usepackage{graphicx,color}
\usepackage{amsmath}
\usepackage{amssymb}
\pdfoutput=1
\usepackage{multirow}
\usepackage{url}
\usepackage{subfigure}
\usepackage{authblk}
\usepackage{fullpage}
\usepackage{lscape}

\begin{document}

\title{On-site Background Measurements for the J-PARC E56 Experiment: A Search for Sterile Neutrino at J-PARC MLF}
\author[1]{S.~Ajimura}
\author[2]{T.~J.~C.~Bezerra}
\author[2]{E.~Chauveau}
\author[2]{T.~Enomoto}
\author[2]{H.~Furuta}
\author[3]{M.~Harada}
\author[3]{S.~Hasegawa}
\author[1]{T.~Hiraiwa}
\author[4]{Y.~Igarashi}
\author[4]{E.~Iwai}
\author[4]{T.~Maruyama}
\author[3]{S.~Meigo}
\author[1]{T.~Nakano}
\author[5]{M.~Niiyama}
\author[4]{K.~Nishikawa}
\author[1]{M.~Nomachi}
\author[4]{R.~Ohta\footnote{Now at Quintessa K.K., Kanagawa, JAPAN}}
\author[2]{H.~Sakai}
\author[3]{K.~Sakai}
\author[3]{S.~Sakamoto}
\author[1]{T.~Shima}
\author[2]{F.~Suekane}
\author[4]{S.~Y.~Suzuki}
\author[3]{K.~Suzuya}
\author[4]{K.~Tauchi}

\affil[1]{\it{\small {Research Center for Nuclear Physics, Osaka University, Osaka, JAPAN}}}
\affil[2]{\it{\small {Research Center for Neutrino Science, Tohoku University, Sendai, Miyagi, JAPAN}}}
\affil[3]{\it{\small {J-PARC Center, JAEA, Tokai, Ibaraki JAPAN}}}
\affil[4]{\it{\small {High Energy Accelerator Research Organization (KEK), Tsukuba, Ibaraki, JAPAN}}}
\affil[5]{\it{\small {Department of Physics, Kyoto University, Kyoto, JAPAN}}}

\date{\today}
\maketitle

\abstract{
The J-PARC E56 experiment aims to search for sterile neutrinos at the J-PARC
Materials and Life Science Experimental Facility (MLF). 
In order to examine the feasibility of
the experiment, we measured the background rates of different 
detector candidate sites,
which are located at the third floor of the MLF, using a detector 
consisting of plastic scintillators with a fiducial mass of 500~kg.
The gammas and neutrons induced by the beam as well as the backgrounds from the
cosmic rays were measured, and the results are described in this article. 
}

\normalsize
\thispagestyle{empty}

\newpage

\setcounter{figure}{0}
\setcounter{table}{0}
\indent

\section{Introduction}
\indent

\begin{figure}[h]
 \centering
 \includegraphics[width=1.1 \textwidth]{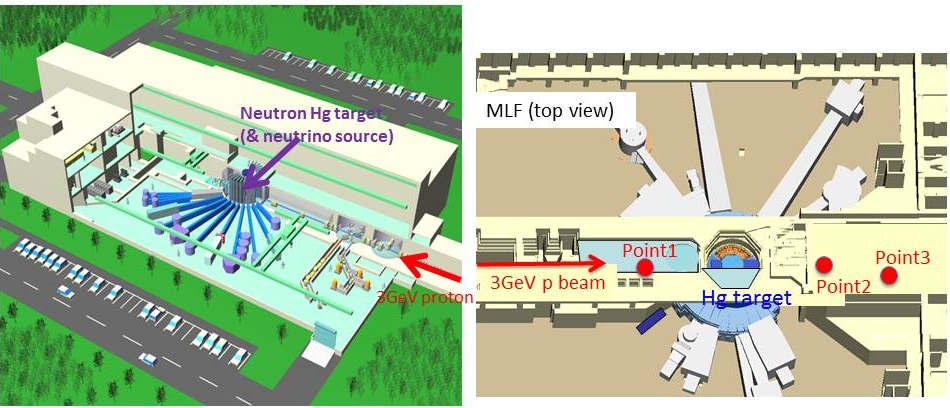}
\caption{
A bird eyes' view of the the MLF facility in J-PARC (left), and the
measurement locations (red circles) of the experiment. (right; 
those are written as ``Point~1'', ``Point~2'' and ``Point~3''))
}
 \label{FIG:MLF}
\end{figure}

We proposed a definite search for the existence of neutrino oscillations
with $\Delta m^2$ near 1~eV$^2$ at the J-PARC Materials and Life Science
Experimental Facility (MLF)~\cite{CITE:P56Proposal}.  
With the 3 GeV Rapid Cycling Synchrotron 
(RCS) and spallation neutron target, an intense neutrino beam from muon
decay at rest ($\mu DAR$) is available. Neutrinos come predominantly from the
$\mu^+$ decay : $\mu^+\rightarrow e^+ +\bar{\nu}_{\mu} +\nu_e$.
The oscillation to be searched for is 
$\bar{\nu}_{\mu} \rightarrow \bar{\nu}_e$  
which is detected by the inverse $\beta$ decay (IBD) 
interaction $\bar{\nu}_e+p \rightarrow e^++ n$, followed by gammas from
neutron capture. 
Figure~\ref{FIG:MLF} (left) shows a bird eyes' view of the 
MLF building.

The unique features of the proposed experiment, compared with the prior
LSND experiment~\cite{CITE:LSND} and 
experiments using conventional horn focused beams (e.g.:~\cite{CITE:MB}), are;

\begin{enumerate}
\item The pulsed proton beam with about 600~ns spill width from J-PARC RCS
and muon long lifetime allows us to select neutrinos from $\mu DAR$ 
(The protons are produced with a repetition rate of 25 Hz, where each 
spill contains two 100 ns wide pulses of protons spaced 600 ns apart).
This can be easily achieved by gating out for about 1~$\mu$s from the start
of the proton beam spill, that eliminates neutrinos from pion and kaon
decay-in-flight.
\item The time gate width with $\sim$10~$\mu$s to obtain the neutrinos
from $\mu DAR$ can also provide superb rejection capability on the cosmic 
ray background by a factor of $\sim$ 4000. 
\item The spallation neutron source is the mercury target, which is 
high-$Z$ material, surrounded by thick iron and concrete shields as shown
in Fig.~\ref{FIG:TARGET}.  
Due to a strong nuclear absorption of $\pi^-$ and $\mu^-$ in the mercury target, neutrinos from $\mu^-$ decay are strongly suppressed up to about the $10^{-3}$ level. The resulting neutrino beam is predominantly $\nu_{e}$ and $\bar{\nu}_{\mu}$ from $\mu^{+}$ with contamination from other neutrino species at the level of $10^{-3}$. 
\begin{figure}[htbp]
\centering
\subfigure[The mercury target]{
\includegraphics[width=0.45\textwidth,angle=0]{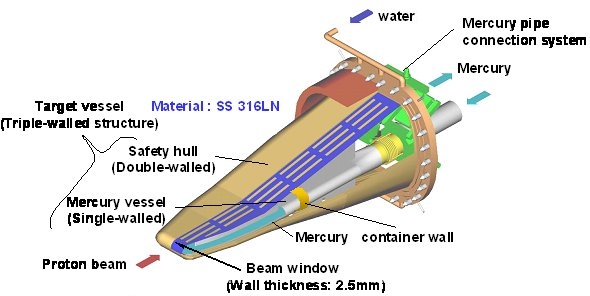}
}
\subfigure[Structure around the mercury target]{
\includegraphics[width=0.45\textwidth,angle=0]{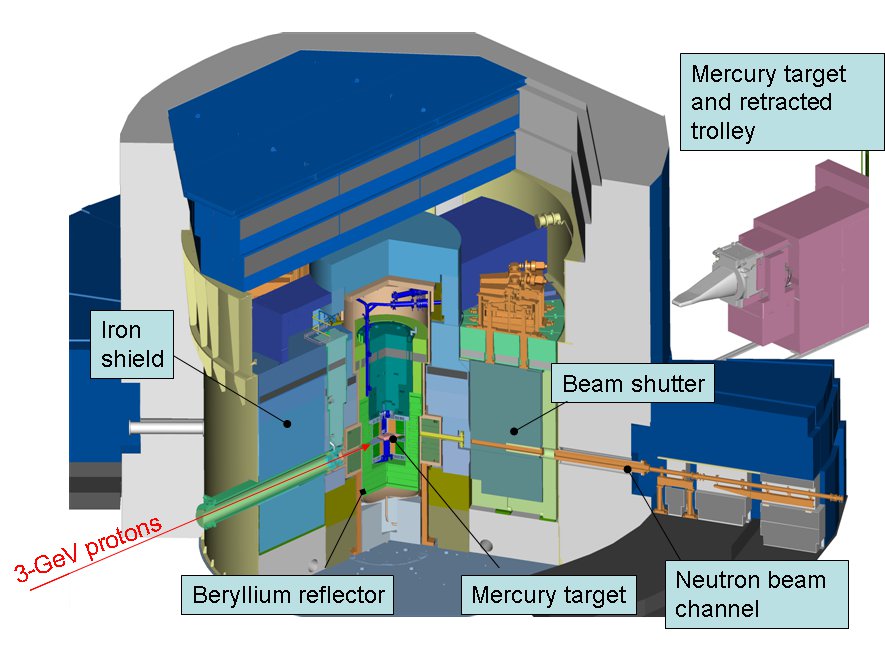}
}
\caption{Mercury target (left) and the structure, which surrounds the target (right).
} 
\label{FIG:TARGET}
\end{figure}
\item $\bar{\nu}_e$ interacts via IBD and its cross section is known to a few percent accuracy~\cite{CITE:IBD}.
\item The neutrino energy can be reconstructed from the positron visible 
energy by adding $\sim$ 0.8~ MeV. 
\item The $\bar{\nu}_{\mu}$ and $\nu_{e}$ fluxes have different and well defined
spectra. This allows us to separate $\bar{\nu}_e$ due to 
$\bar{\nu}_{\mu} \rightarrow \bar{\nu}_e$ oscillations from those due to 
$\mu^-$ decay contamination.
\end{enumerate}

\begin{figure}[h]
 \centering
 \includegraphics[width=1.1 \textwidth]{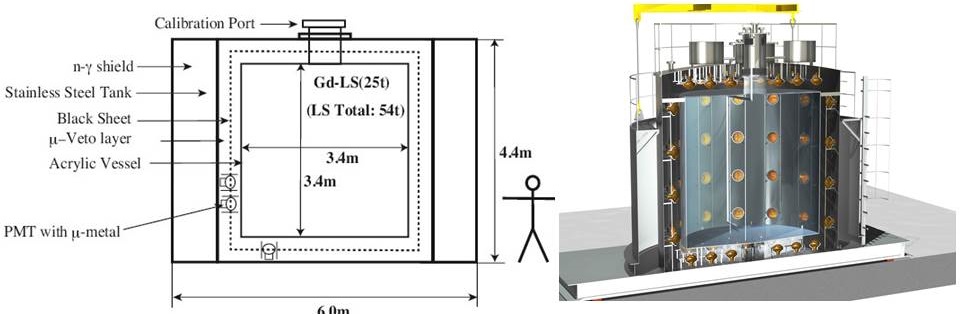}
\caption{
Current design of the E56 25 tons detector. Left: schematic view; right:
3D drawings. 
}
 \label{FIG:50ton}
\end{figure}

In addition to these, Gd-loaded liquid scintillator~\cite{GdLS} is used to 
reduce the accidental backgrounds in the E56 experiment.
Figure~\ref{FIG:50ton} shows the current detector design.

In the proposal, two detectors are placed on the third floor of MLF with a 
baseline $\sim$20~meters from the mercury target. 

In order to examine the feasibility of the J-PARC E56 experiment,  
we carried out an on-site test experiment (MLF 2014BU1301 experiment), 
which was mainly dedicated to measure the beam related backgrounds.
The required accuracy of the measurements is a few ten $\%$ to check the
feasibility of the E56.
The data was taken from April to July 2014 using a 500~kg plastic scintillator.
The setup and calibration of the detector, and  
the results of background rate measurements are described in this article.

\section{Definition of the IBD Signal}
\indent

Before describing the results, the definition of the IBD signal inside
the liquid scintillator is clarified since it will be used often within the text.
As explained before, the IBD interaction is  
$\bar{\nu}_e+p \rightarrow e^++ n$, followed by gammas from the
neutron capture. For the detection of the positron, which we call ``prompt 
signal region'', the energy is selected to be 20$< E <$60 MeV and the
hit time of the activity inside the detector to be 1$< T <$10 $\mu$s, where
$T$ is the time from the proton beam start timing. 
These criteria are based on the features of the $\mu DAR$ neutrinos. 
On the other hand, to catch the delayed gamma 
(called ``delayed signal region''), 
we select the activity, which have the energy in the range
7$< E <$12 MeV and the hit time to be $<$ 100~$\mu$s.    

To examine the background rate, these selection criteria are used unless
noted otherwise.

\section{Plastic Scintillators with 500 kg Fiducial Volume (the 500kg Detector)}
\indent

The 500 kg plastic scintillator detector consists of the main target scintillators 
and two layers of charged particle vetoing system.
We describe the setup, the calibration and the obtained resolution of 
the scintillator detector in the following subsections.

\subsection{Setup}
\label{appendix_500kg_setup}
\indent

\begin{figure}[hbtp]
	\begin{center}
		\includegraphics[scale=0.42]{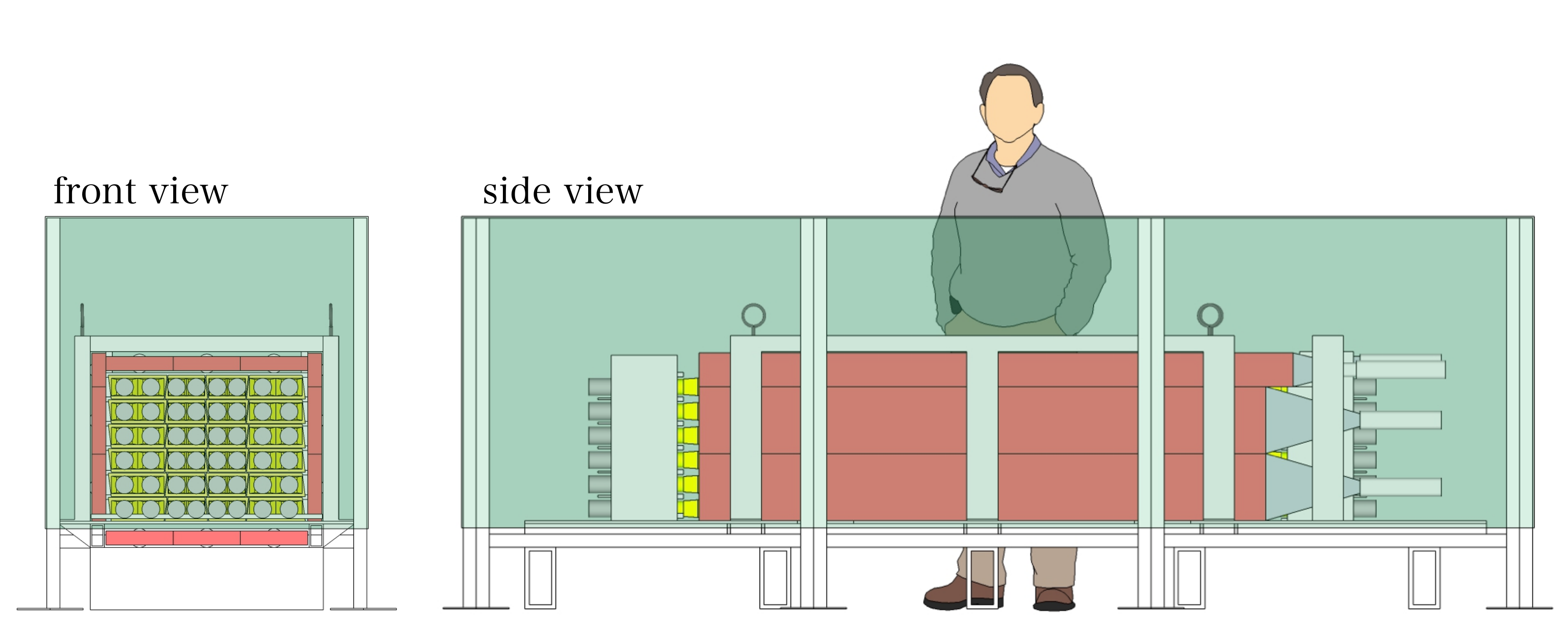}
		\caption{Schematic view of the 500 kg plastic scintillator detector at the third floor of MLF(left: front view, right: side view).
		The 500 kg plastic scintillators (yellow) were surrounded by two layers of charged particle vetoing system, Inner veto(red) and Outer veto(green).}
		\label{500kg_setup}
	\end{center}
\end{figure}

Figure~\ref{500kg_setup} shows a schematic view of the 500 kg plastic scintillator counters placed at the third floor of MLF. The yellow parts of the
Fig.~\ref{500kg_setup} show the 500~kg main scintillators in this experiment.   
They consist of two types of scintillators: 12 pieces of $11.7/13.7$ (trapezoid) $\times 7.6 \times 182$~cm$^3$ scintillator (1D)
and 12 pieces of $16.9/18.8$ (trapezoid) $\times 7.6 \times\ 182$~cm$^3$ scintillator (3D). To minimize the dead space, 3D scintillators were located on the both sides, while 1D scintillators were located in the central part.
Each end of the scintillators was viewed by two PMTs. 
Signals from each PMT were recorded by FINESSE 500 MHz 8 bit 
FADC~\cite{CITE:FINESSE}.
The 500 kg detector was surrounded by two layers of charged particle
vetoing systems, the Inner and Outer vetoes.
The Inner veto (red parts in Fig.~\ref{500kg_setup}) 
covers the surfaces of the top, bottom and both 
sides of the 500 kg detector.
The thickness of the plastic scintillator for Inner veto is 4.3 cm.
The Outer veto (green parts in Fig.~\ref{500kg_setup}) 
surrounds the 500 kg detector and Inner veto, with mostly 
2 layers of 6-8~mm thick plastic scintillators.
PMT signals from the veto counters were recorded by FINESSE 65 MHz FADC 
with 50 ns RC-filter~\cite{CITE:FINESSE}.

\subsection{Calibration}
\indent

\begin{figure}[hbtp]
	\begin{center}
		\includegraphics[scale=0.6]{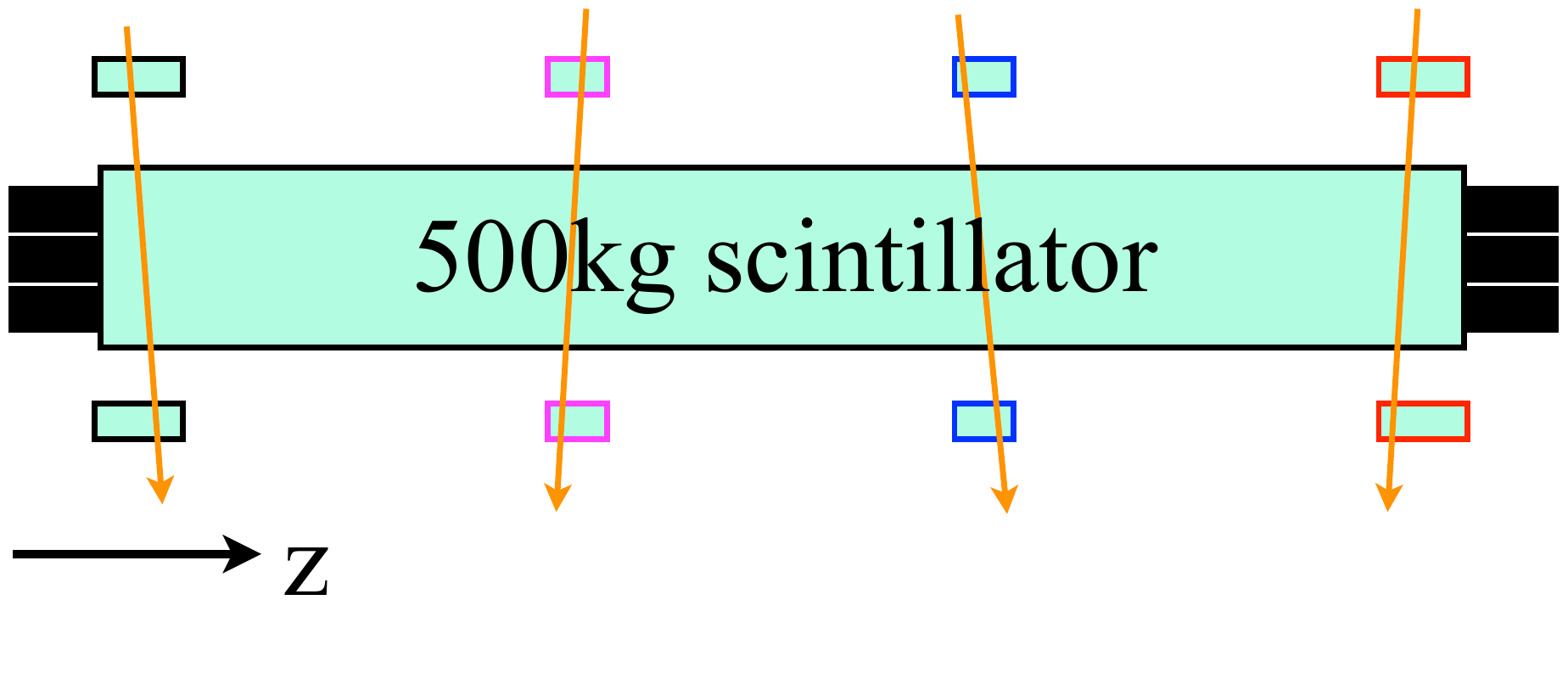}
		\caption{Schematic view of the cosmic muon trigger counters.
		Four pairs of plastic scintillators were prepared to trigger cosmic muons.}
		\label{500kg_setup2}
	\end{center}
\end{figure}

We used cosmic muons to calibrate the energy and timing and to measure the attenuation length of the scintillator.
Four pairs of plastic scintillators were prepared to define the cosmic 
ray tracks.
Figure~\ref{500kg_setup2} shows a schematic view of the cosmic muon trigger counters. The size of the scintillators located at both sides are 132~cm (length) $\times$ 10.5~cm (width) $\times$ 4.0~cm (thickness), while those located in the 
center are typically 80~cm (length) $\times$ 5.0~cm (width) $\times$ 1.5~cm (thickness).

\begin{figure}[hbtp]
	\begin{center}
		\includegraphics[scale=0.4]{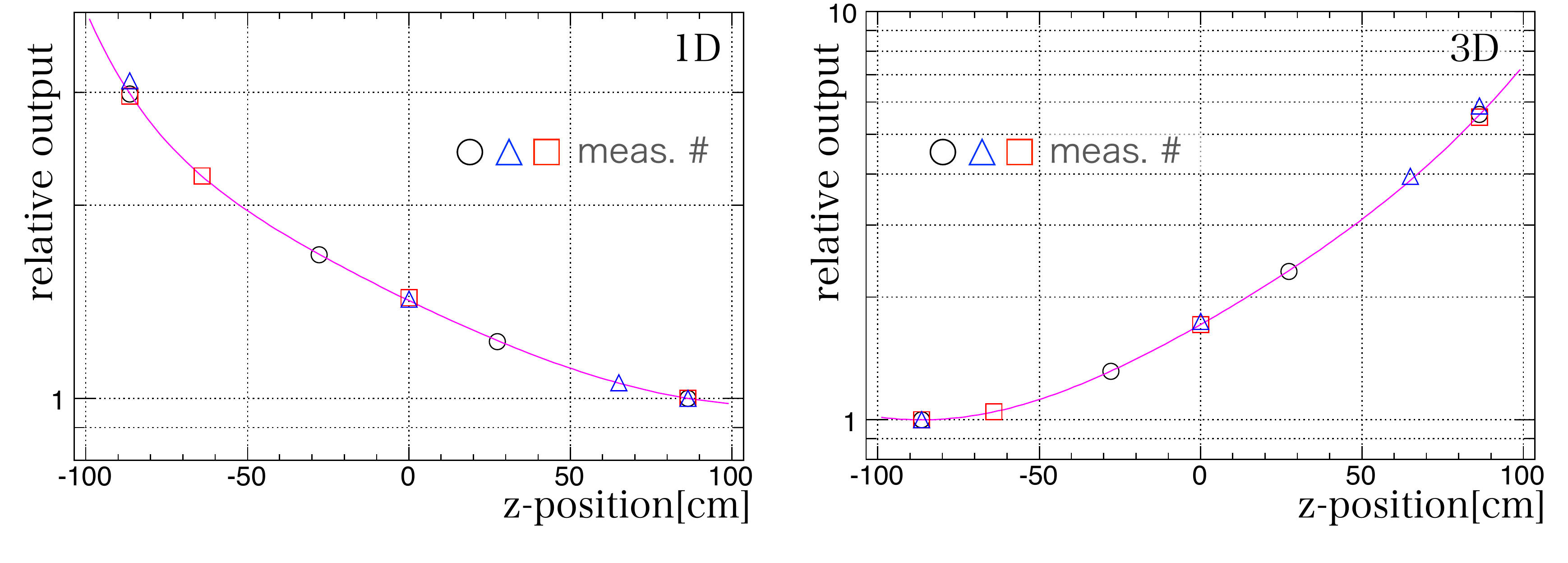}
		\caption{Typical attenuation curve for each scintillation type (1D and 3D).
		We measured the curves and parametrized for each scintillator (magenta line. The fit function is the 2nd polynomial + exponential).}
		\label{500kg_attenuation}
	\end{center}
\end{figure}

To measure the attenuation length of the scintillators, we made some dedicated runs in which we changed the position of the trigger counters.
Figure~\ref{500kg_attenuation} shows the typical attenuation curves measured for each scintillator type.
We measured the attenuation curve and parametrized for each scintillator.
By considering the attenuation length, the reconstructed charge becomes 
independent from the incident position.

The gain of each PMT was also calibrated by using cosmic ray events.
The output from each scintillator is stable within 1-2\% during the
measurements with the gain correction.

We also adjusted the timing of each PMT.
Time offsets were determined to minimize the time difference between each pair of PMTs on each end.
The velocity of cosmic muons passing through the detector was considered, where
the typical light velocity inside the scintillators is 14.3 cm/ns.

\subsection{Resolutions}
\indent

We also evaluated the position and energy resolutions of the 500 kg detector using
the cosmic ray events. As the results of the precise timing calibrations, 
the obtained position resolution is $\sigma_z=2.6$ cm for MIP energy.
The typical energy resolutions of the scintillators are 3.3\% for 1D 
and 4.5\% for 3D at the middle of the scintillators for MIP energy, which
is about 13 MeV.

\subsection{Detector simulation}
\label{sc_500kg_mc}
The resolutions described above and other detector responses, such as quenching effects with Birks' law, light attenuation and the threshold effect, were implemented to Geant4~\cite{GEANT4} based Monte Carlo simulation (MC).

\subsection{Veto Efficiency}
\label{sec:VetoEff}
\indent

In order to measure the efficiency of the Inner Veto (IV) and Outer Veto (OV) systems on the background measurement, their particle tagging efficiency was measured. The veto efficiency ($\varepsilon$) is defined as follows:

\begin{eqnarray*}
\varepsilon_{\rm IV} & = & \frac{\rm N\ of\ coincident\ event\ OV,\ IV\ and\ Target\ (Triple\ Coincidence)}{\rm N\ of\ coincident\ event\ OV\ and\ Target\ (Double\ Coincidence)} \\
\nonumber \\
\varepsilon_{\rm OV} & = & \frac{\rm N\ of\ coincident\ event\ OV,\ IV\ and\ Target\ (Triple\ Coincidence)}{\rm N\ of\ coincident\ event\ IV\ and\ Target\ (Double\ Coincidence)}
\end{eqnarray*}

\begin{figure}[htpb!]
	\centering \includegraphics[width=0.75\textwidth]{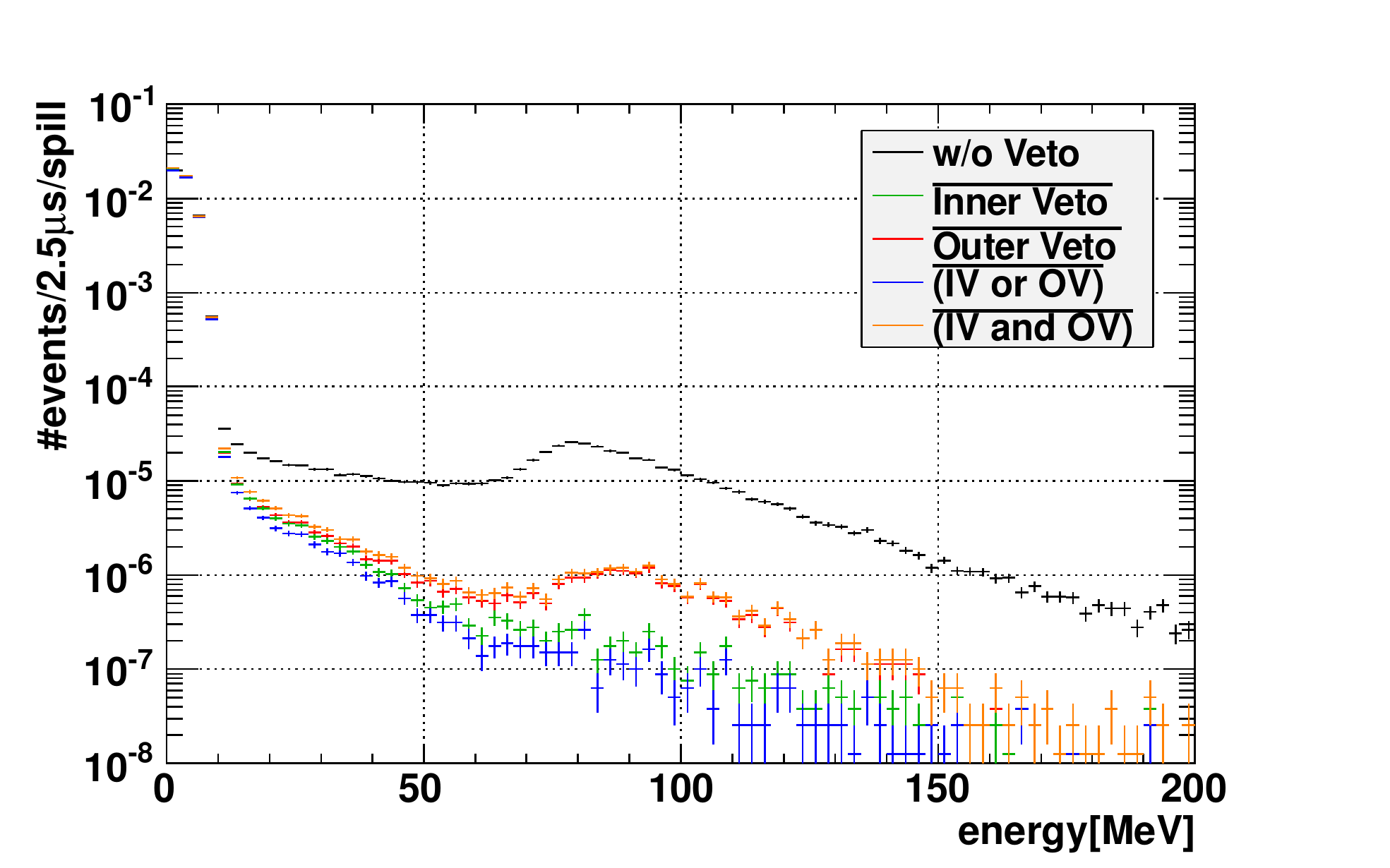}
	\caption{Energy spectra for the 500 kg detector for different veto conditions. The peak due to the cosmic muons are seen around 80 MeV for the distribution without veto.}
	\label{fig:VetoEff}
\end{figure}

The energy spectra of the main 500~kg detector for different veto conditions 
in no beam period are displayed in Fig~\ref{fig:VetoEff}. 
Around the 80 MeV region of the energy distribution without veto, there is a peak 
created by the cosmic muons. 
Finally, the veto efficiency for different energy ranges is summarized in Table~\ref{tab:VetoEff_sum}, where it can be seen that a total veto efficiency is better than 99.8\% assuming the IV and OV inefficiencies are independent. In these Table and Figure, it is also possible to see that muons deposit an energy larger than 60~MeV and most of cosmic ray muon background are thus rejected.

\begin{table}[htpb!]
	\caption{Summary of IV and OV efficiency for different energy ranges.}
	\begin{center}
		\begin{tabular}{cccc}
			\hline \hline
			Energy Range [MeV]	&	$\rm \varepsilon_{IV}$	&	$\rm \varepsilon_{OV}$	&	(IV or OV)		\\
			\hline
			20 $<$ E $<$ 60		&	96.8$\pm$0.2\%			&	94.1$\pm$0.2\%			&	$\sim$99.8\%		\\
			60 $<$ E $<$ 100		&	99.5$\pm$0.04\%			&	96.2$\pm$0.1\%			&	$\sim$99.9\%		\\
			100 $<$ E $<$140		&	99.6$\pm$0.07\%			&	95.1$\pm$0.3\%			&	$\sim$99.9\%		\\
			\hline \hline
		\end{tabular}
	\end{center}
	\label{tab:VetoEff_sum}
\end{table}

\section{Background Measurement Locations}
\indent

Figure~\ref{FIG:MLF} shows the overview of the MLF building (left), and the 
measurement locations (right). The 500~kg 
detector was moved to measure the backgrounds at each point.  
The baselines from the
mercury target are $\sim$17~m, $\sim$20~m, $\sim$34~m for 
``Point~1'', ``2'' and ``3'' in Fig.~\ref{FIG:MLF},
respectively. We accumulated data for two weeks per each point. 

Only the results of ``Point~2'' and ``Point~3'' are described in this
article. 
The results of ``Point~1'' is expected to be published later since 
the analysis is complicated due to large amount of the neutron background.

The number of spills during the beam to be used for
the analysis are 19,545,739 for ``Point~2'' and 
26,225,816 for ``Point~3'', respectively.

\section{Fast Neutrons Background from Beam}
\label{SEC:ME}
\indent

One of the main purposes of the background measurement is to measure the Michel electron background induced by beam fast neutrons, which was indicated by the previous background measurement at the first floor of MLF 
building~\cite{CITE:P56Proposal}.
According to Geant4, fast neutrons whose kinetic energy are larger than 200 MeV can produce charged pions. The charged pions create the Michel positrons, which has the same energy and hit timing as IBD events in the $\pi \to \mu \to e$ decay-chain and the scattered fast neutrons are thermalized and captured by Gd. 
Therefore, this could be a serious correlated background for the E56 experiment. The flux of such fast neutrons at the third floor of MLF was estimated to be some orders of magnitude smaller than that in the first floor~\cite{CITE:P56Proposal}.
We briefly describe the basic idea and backgrounds of this measurement.

\begin{figure}[hbtp]
	\begin{center}
		\includegraphics[scale=0.38]{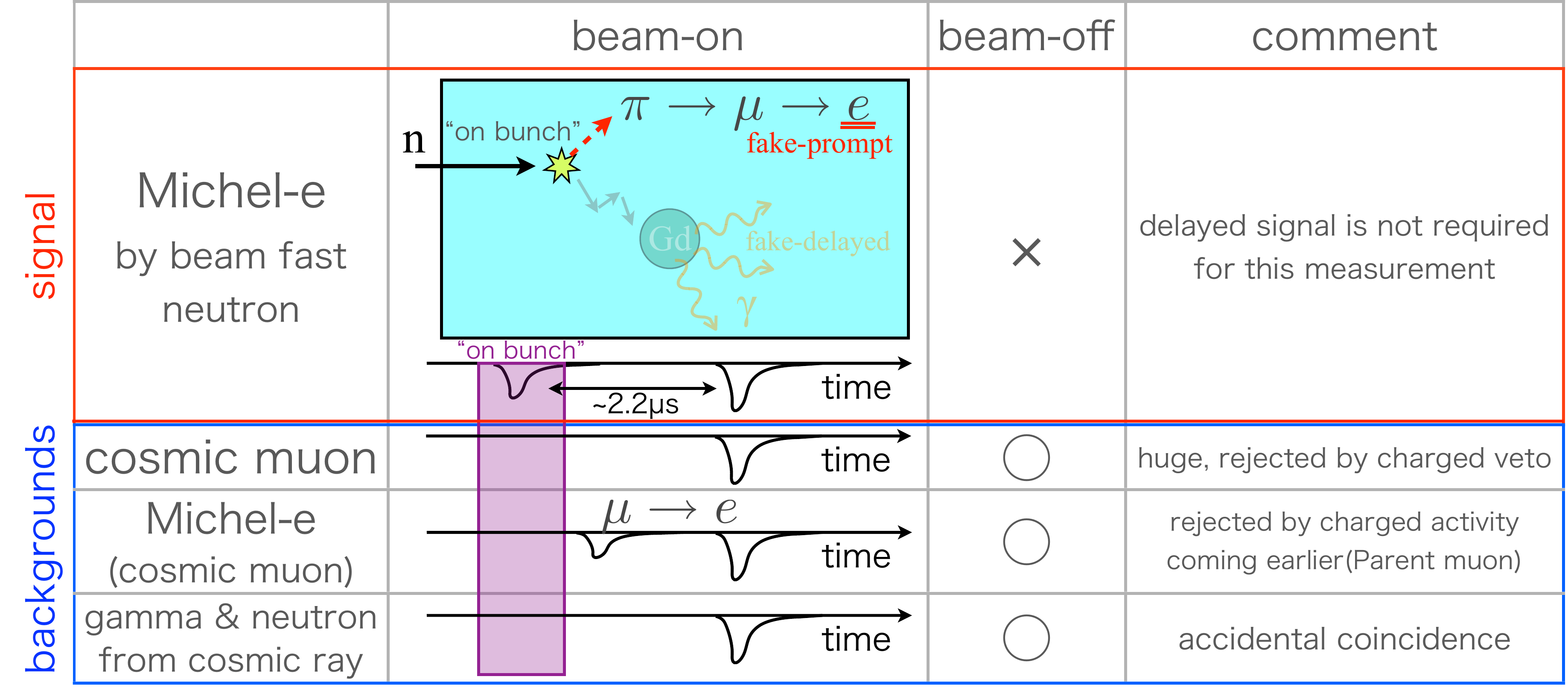}
		\caption{ The definition of ``signal'' and ``backgrounds'' of this measurement.
		Signal is a Michel electron induced by beam fast neutrons.
		Backgrounds are clipping cosmic muons, Michel electrons from cosmic muons and neutral particles from cosmic rays.}
		\label{500kg_measurement}
	\end{center}
\end{figure}

Figure \ref{500kg_measurement} shows the definitions of the ``signal'' 
and ``background'' of this measurement.
The signal is a Michel electron induced by beam fast neutrons.
Fast neutrons coming on the (proton) beam bunch timing hit our detector 
and produce pions.
These pions then produce Michel electrons in the decay chain of  
$n + p$ (or C) $\to  X + \pi^+$, then $\pi^+ \to \mu^+ \to e^+$.
The signature of the signal is thus the coincidence between an activity without any veto hits on the bunch timing and a ``prompt signal" about 2.2 $\mu$s later from the beam timing.
Backgrounds for this measurement are clipping cosmic muons, Michel electrons from cosmic muons and neutral particles (gammas and neutrons) from cosmic rays as shown in Fig.~\ref{500kg_measurement},
and they come either during beam-on or beam-off.
On the other hand, signal comes only when the beam is on.
The basic idea of this measurement is thus to extract signals from backgrounds by subtracting beam-off activities from beam-on activities.

Based on the concept of the measurement, 
we took the following data set:
\begin{itemize}
\item beam-on: To observe activities on the beam bunch and around the bunch timing (the timing definition is described later);
\item beam-off (off-timing) : To subtract backgrounds from beam-on data\\
We took data 20 ms after each beam bunch spill.
Because detector responses such as the PMT gains, efficiency of veto counters and others are exactly the same with the last beam spill,
we can subtract the backgrounds from the beam signals without systematic uncertainties.
\item no-beam: When the accelerator is off. To evaluate truly beam unrelated background;
\item cosmic muons: To calibrate the detector.
\end{itemize}
Note that the radio frequency (RF) of the RCS provides the beam timing signal 
for the measurement, thus we used it to 
take data around the beam and 20~ms later. For the no-beam data, the clock 
trigger was used.

\subsection{Measurement}
\indent

\begin{figure}[htbp]
\centering
\subfigure[``Point~2'']{
\includegraphics[width=0.6\textwidth,angle=0]{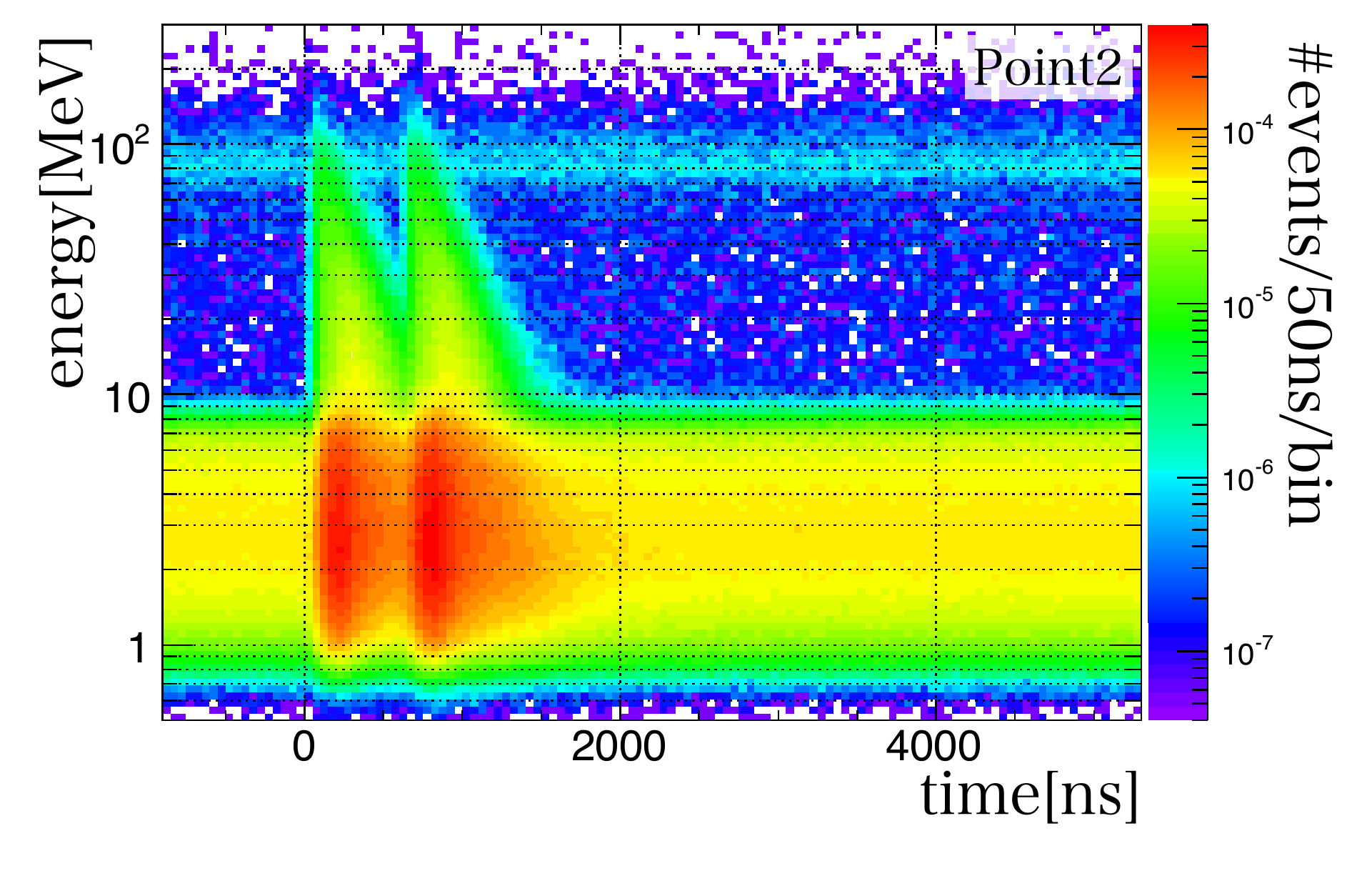}
}
\subfigure[``Point~3'']{
\includegraphics[width=0.6\textwidth,angle=0]{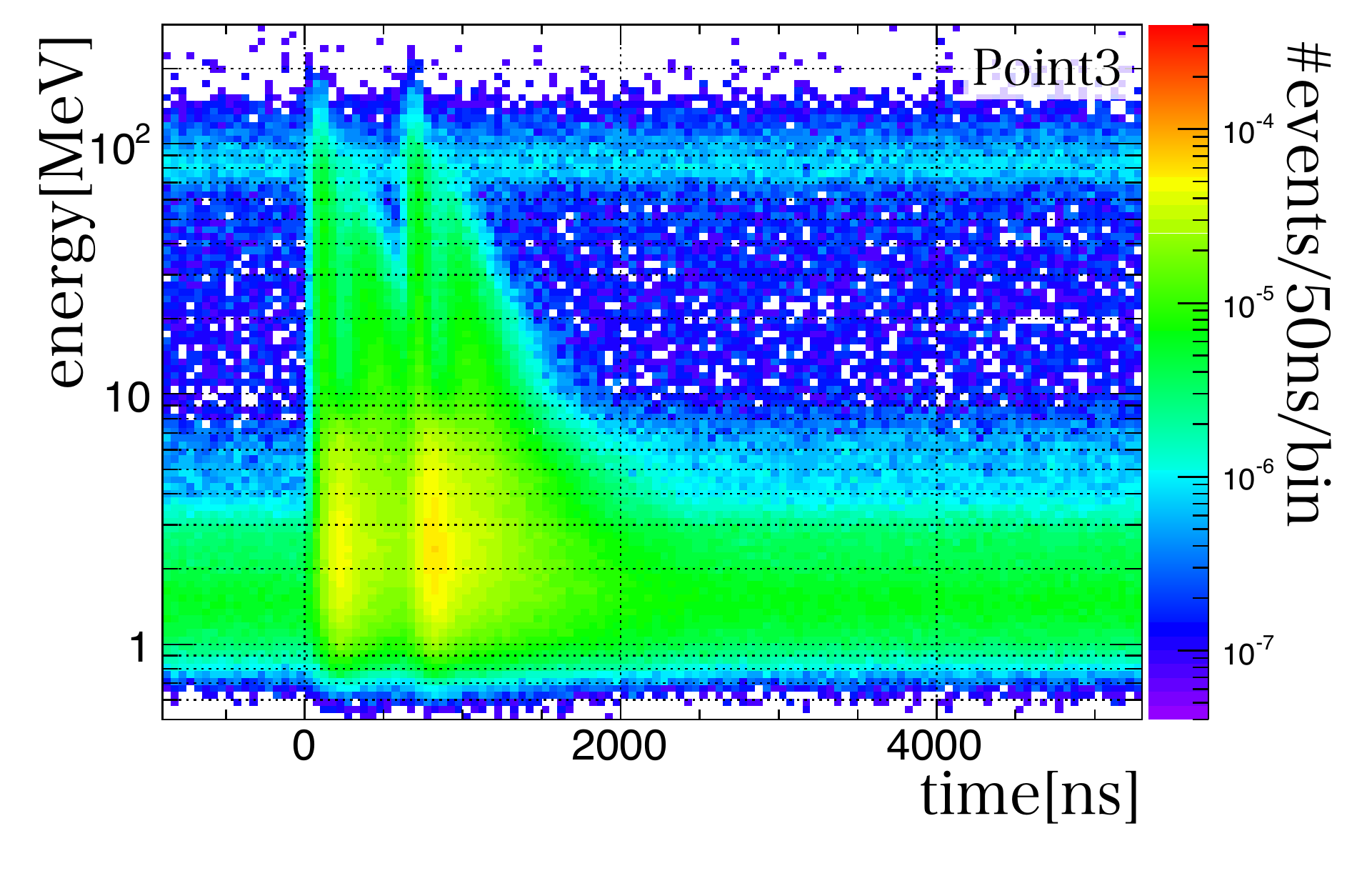}
}
\subfigure[``no-beam'']{
\includegraphics[width=0.6\textwidth,angle=0]{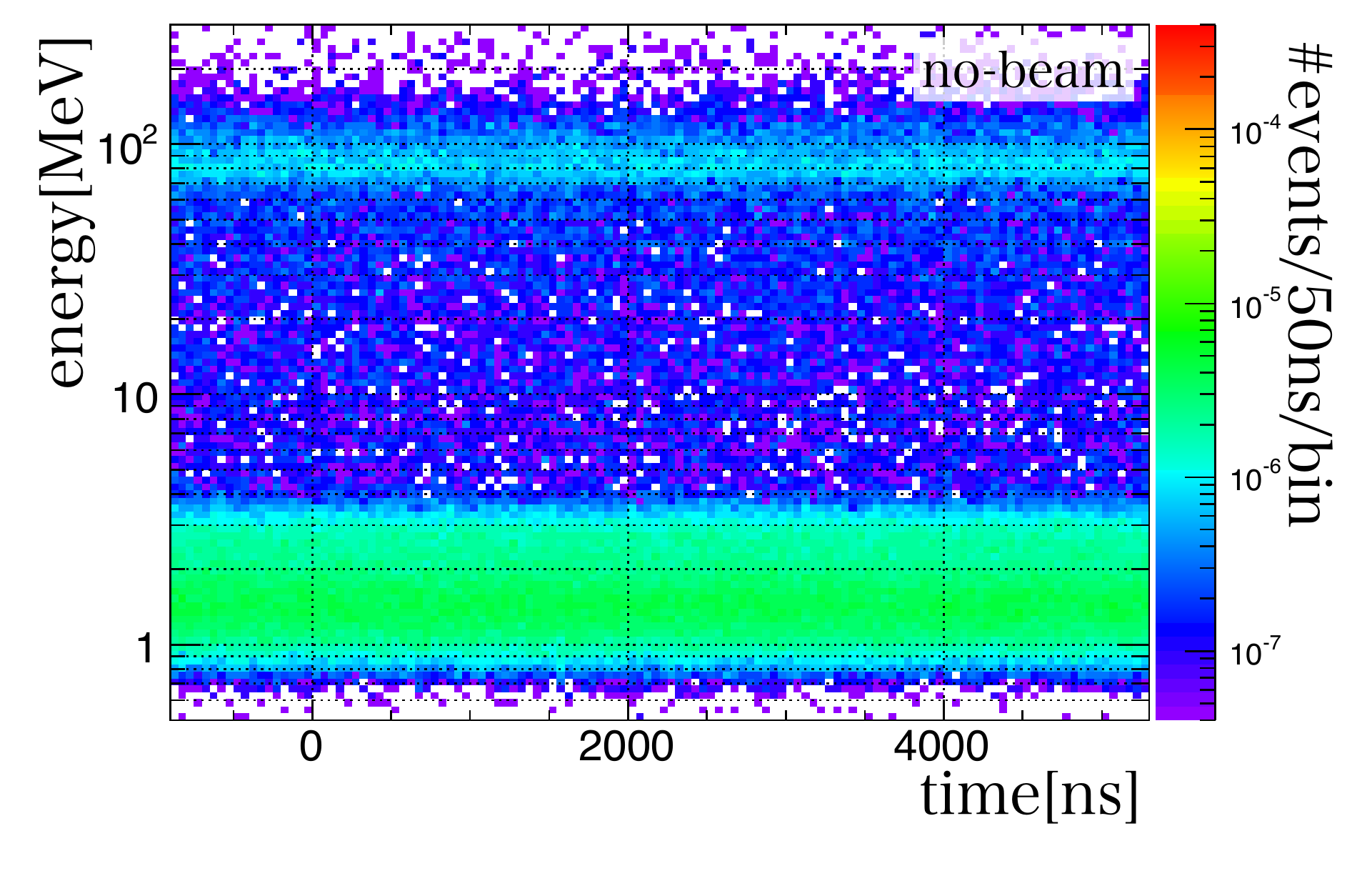}
}
\caption{Correlation between energy and timing of the events observed at (a)``Point~2'', (b)``Point~3'' and (c) ``no-beam''. Note that the vertical axis of the two dimensional plot is based on the log-scale, and one bin of the vertical axis is divided equally in the log-scale.
} 
\label{500kg_TvsE}
\end{figure}

The search for beam neutrons which induce Michel electrons was performed 
by detecting their prompt signals.
Figure~\ref{500kg_TvsE} (a) and (b) show the correlation between the energy and the timing of the events observed at ``Point~2'' and ``Point~3''. The zero
of the horizontal axis corresponds to the starting time of the proton beam.
We observed the large number of activities around the beam time, however
the number of activities are decreased rapidly as a function of time. 
Also no-beam data is shown in Fig.~\ref{500kg_TvsE} (c) for comparison.
Here we used the clock trigger, thus the time axis is arbitrary.  
In this measurement the prompt signal is defined by:
\begin{itemize}
\item $20<E[{\rm MeV}]<60$
\item $1.75<t[{\rm \mu s}]<4.65$ from the rising edge of the first beam bunch.
\end{itemize}
to estimate the number of activities in the IBD prompt region.

\begin{figure}[htbp]
\centering
\subfigure[``Point~2'']{
\includegraphics[width=0.6\textwidth,angle=0]{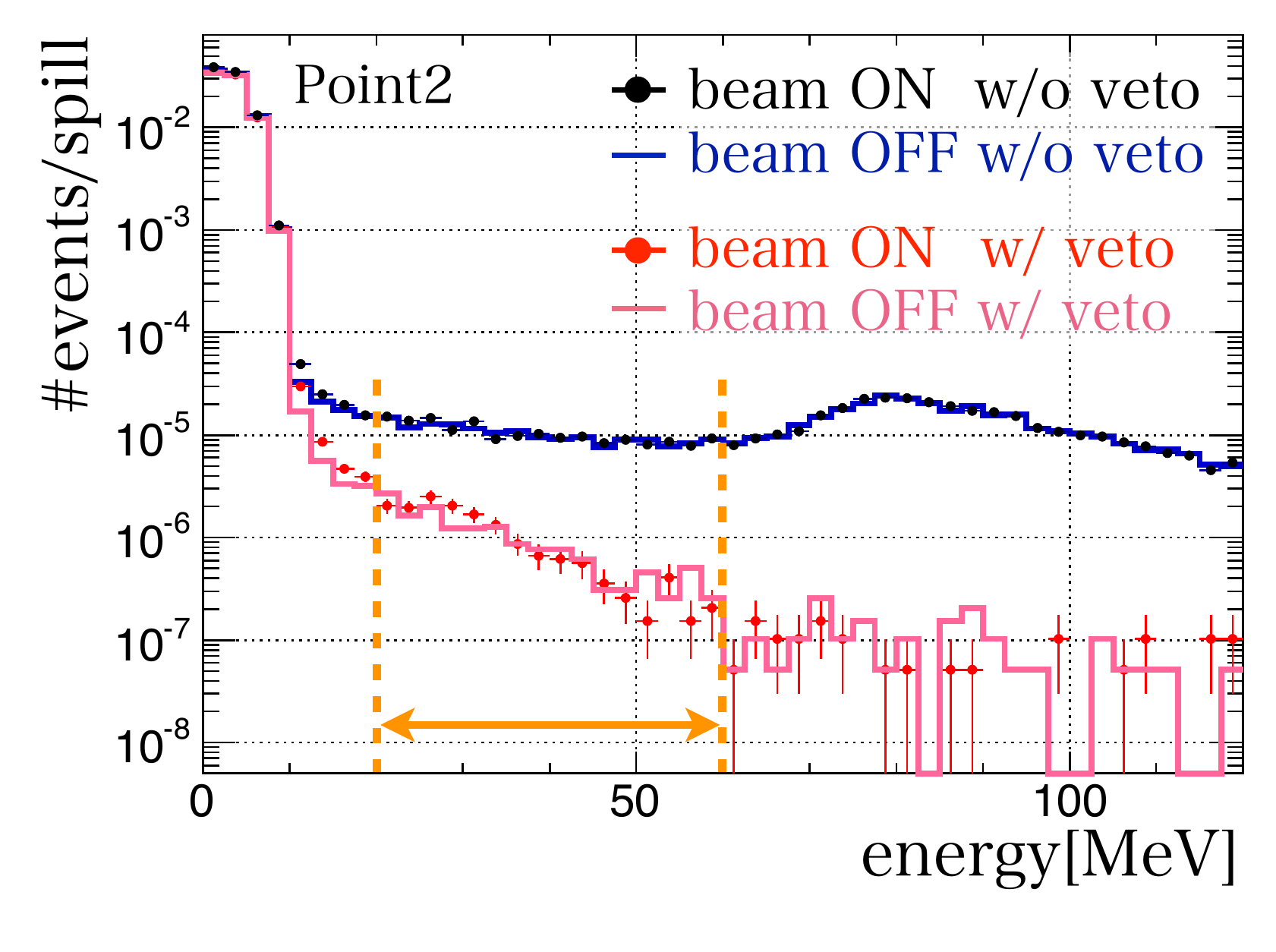}
}
\subfigure[``Point~3'']{
\includegraphics[width=0.6\textwidth,angle=0]{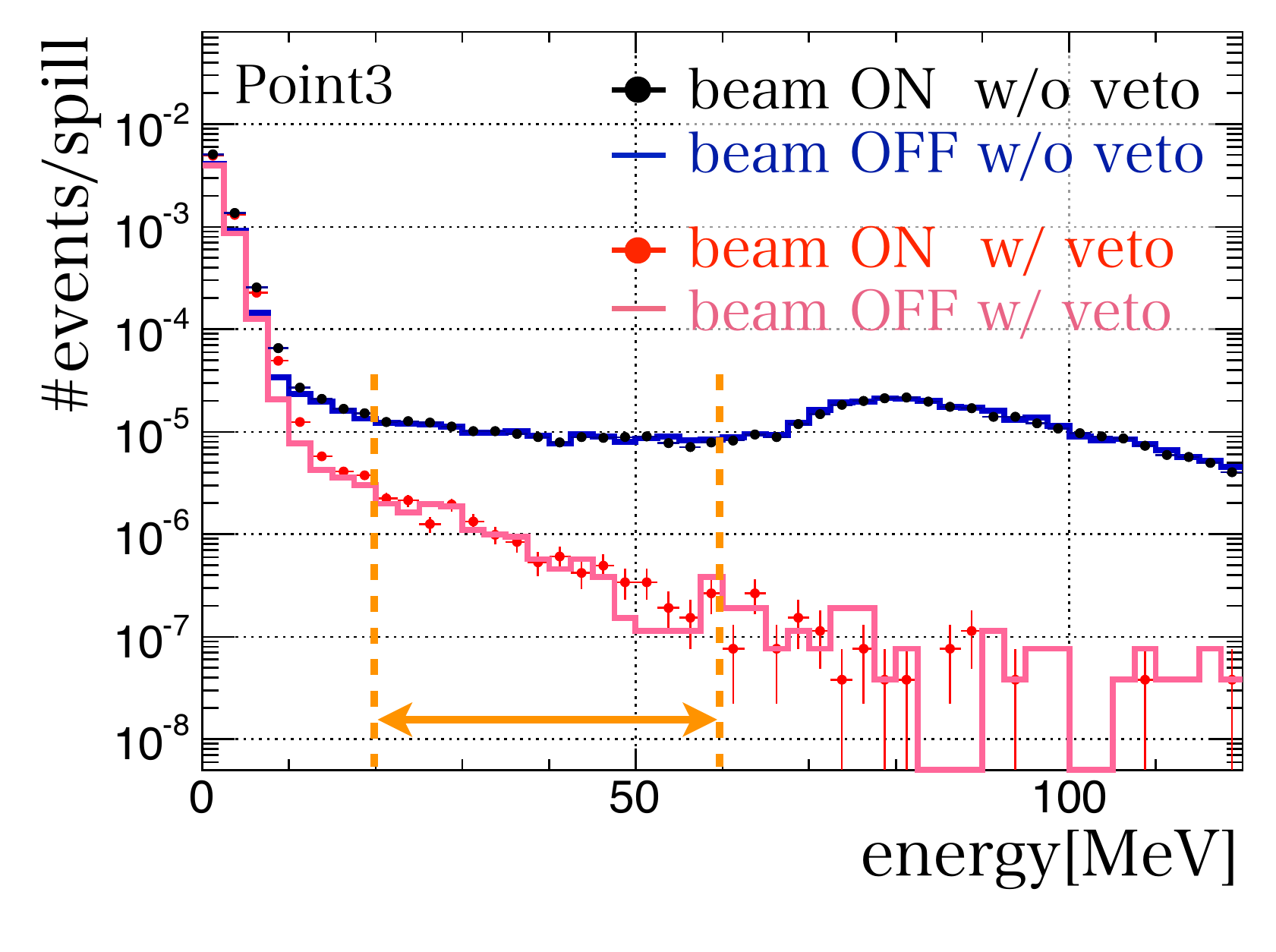}
}
\caption{Energy distributions of events in the prompt timing window and beam-off data, before and after applying the charged veto cut. (a) ``Point~2'' and (b) ``Point~3''.
} 
\label{500kg_michele_edep}
\end{figure}

As described before, we compare the beam-on and beam-off data to subtract other activities without any systematic uncertainties.
A huge number of the clipping muon background (order of a few 100 Hz) was rejected by applying the charged veto cut.
Figures~\ref{500kg_michele_edep} show the energy distributions of events in the prompt timing window, $1.75<t[{\rm \mu s}]<4.65$ from the beam bunches, and beam-off data, before and after applying the charged veto cut. The observed rates 
are summarized in Table~\ref{TAB:prompt_res1}.
The numbers of events in the prompt energy range are both consistent statistically between beam-on and beam-off data either with or without applying the charged veto cut.

\begin{table}[htbp]
\centering
\caption{Summary of the background rates in the prompt region.}
\label{TAB:prompt_res1}
\begin{tabular}{|c|c|c|}
\hline
& ``Point~2'' ($\times 10^{-5}$/spill) & ``Point~3'' ($\times 10^{-5}$/spill)\\
\hline\hline
beam-on w/o veto &  $16.8 \pm 0.3$ & $15.3 \pm 0.2$ \\ 
beam-off w/o veto&  $16.4 \pm 0.3$ & $15.4 \pm 0.2$ \\ \hline
subtraction (on-off) & 0.4$\pm$0.4 & -0.1$\pm$0.3\\ \hline\hline
beam-on w/ veto  & $1.58 \pm 0.09$ & $1.41 \pm 0.07$ \\ 
beam-off w veto  & $1.52 \pm 0.09$ & $1.33 \pm 0.07$ \\ \hline 
subtraction (on-off) & 0.06$\pm$0.13 & 0.08$\pm$0.10\\ 
\hline
\end{tabular}
\end{table}

\begin{figure}[hbtp]
	\begin{center}
		\includegraphics[scale=0.55]{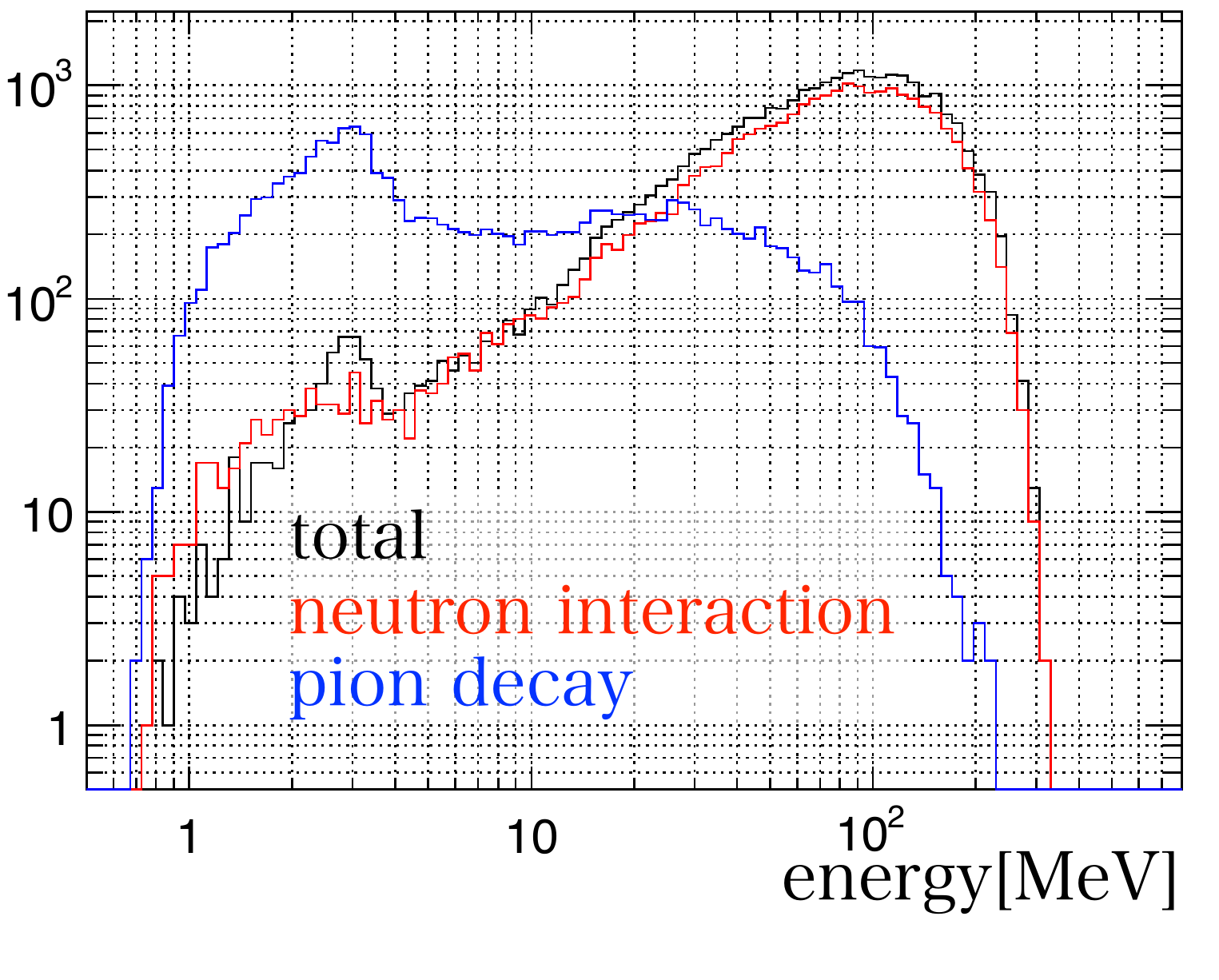}
		\caption{Energy deposit on the bunch timing for beam Michel electron with the neutron kinetic energy 300 to 500 MeV (flat). Note this is MC.
		Red, blue and black lines are the energy deposit from neutron interactions, pion decays and their sum in each event. Note that one bin of the horizontal axis is divided equally in the log-scale, and the unit of the vertical axis is arbitrary.
		A peak around 3 MeV in the blue histogram corresponds to the muon kinetic energy, 4.2 MeV, from stopped pion decay including the quenching effects with the Birks' law.
		The quenching effects, light attenuation in the scintillator and other detector responses such as resolutions and threshold effects are implemented to the Geant4 as described in Section~\ref{sc_500kg_mc}.}
		\label{500kg_onbunch_edep}
	\end{center}
\end{figure}

\begin{figure}[hbtp]
	\begin{center}
		\includegraphics[scale=0.55]{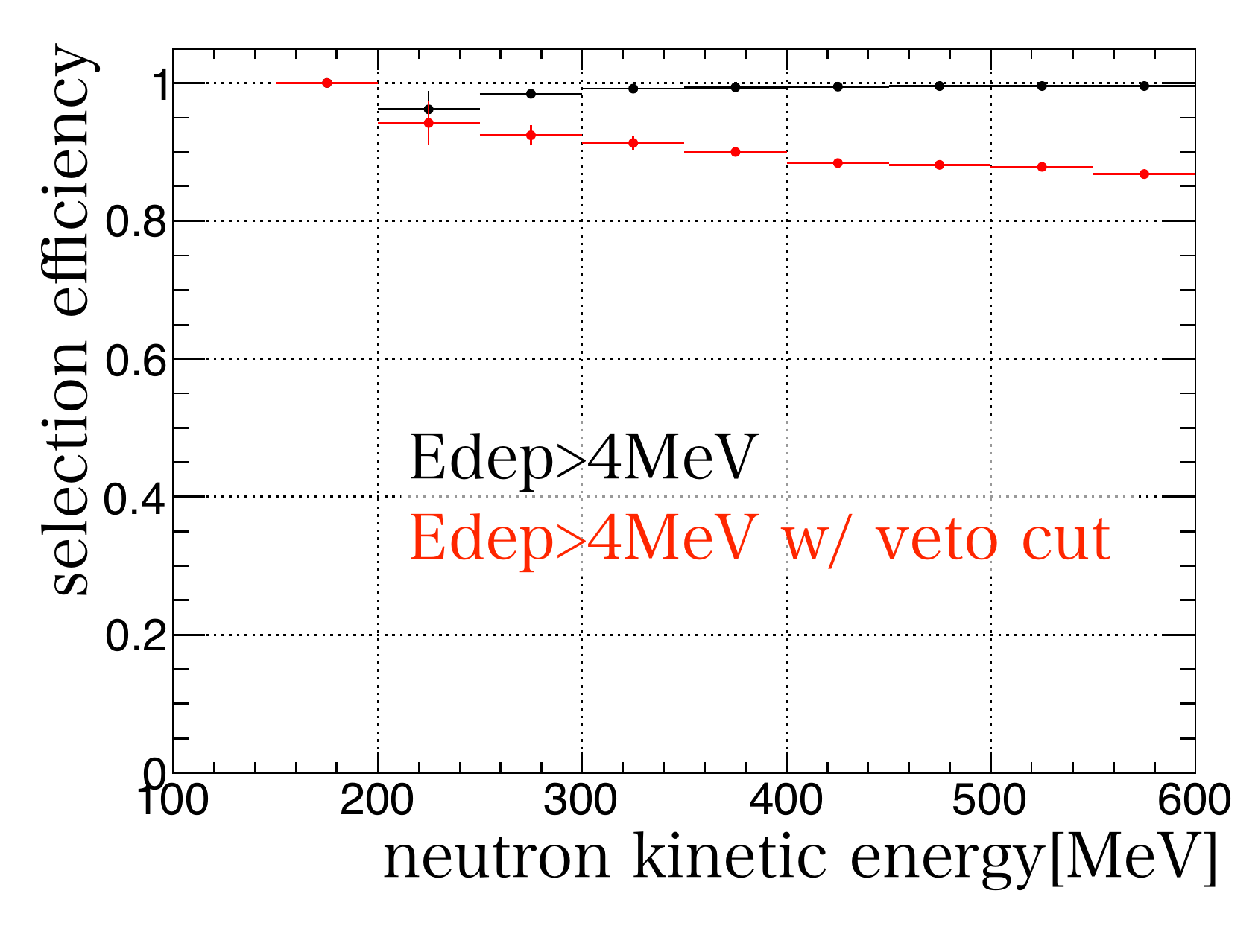}
		\caption{Estimated selection efficiency of the on-bunch cut as a function of the incident neutron kinetic energy.}
		\label{500kg_onbunch_eff}
	\end{center}
\end{figure}

To improve the sensitivity, an additional cut was applied before obtaining 
the final result.
Figure~\ref{500kg_onbunch_edep} shows a simulated energy distribution on the bunch timing when fast neutrons produce charged pions in the 500 kg detector.
We can expect some events on the bunch timing associated with beam Michel electron backgrounds.
On the other hand, as shown in Fig.~\ref{500kg_measurement},  the beam unrelated backgrounds have no activities on the bunch timing.
Michel electrons from cosmic muons can have an activity on the bunch timing only when the parent muon comes on the bunch timing accidentally.
However, since the muon is a charged particle, it can be easily rejected by the veto counters.
We can thus strongly suppress backgrounds by requiring on-bunch activities without hits in the veto counters, and
at least one on-bunch activity (Edep$>$4 MeV) with this veto conditions was required.
Figure \ref{500kg_onbunch_eff} shows the estimated selection efficiency of this on-bunch cut as a function of the incident neutron kinetic energy based on MC.
Though most of the events have more than 4 MeV energy deposit at on the bunch timing, a part of the events are rejected by self-vetoing.
The selection efficiency has slight dependence on the incident neutron kinetic energy.
Because we do not know the energy spectrum of the incident neutron well, we assumed the selection efficiency, $\epsilon_{\rm on bunch}=0.9$.

\begin{figure}[htbp]
\centering
\subfigure[``Point~2'']{
\includegraphics[width=0.6\textwidth,angle=0]{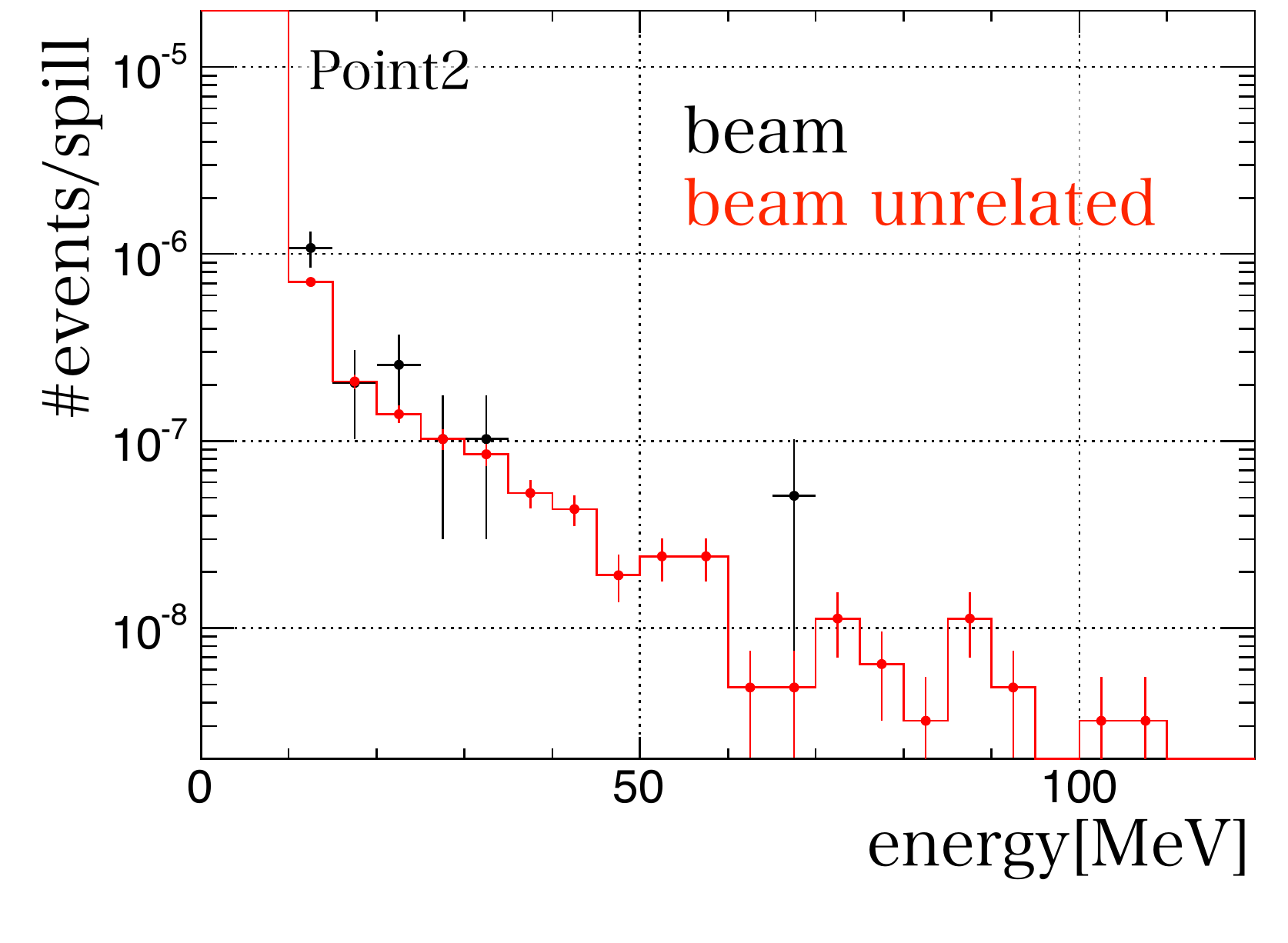}
}
\subfigure[``Point~3'']{
\includegraphics[width=0.6\textwidth,angle=0]{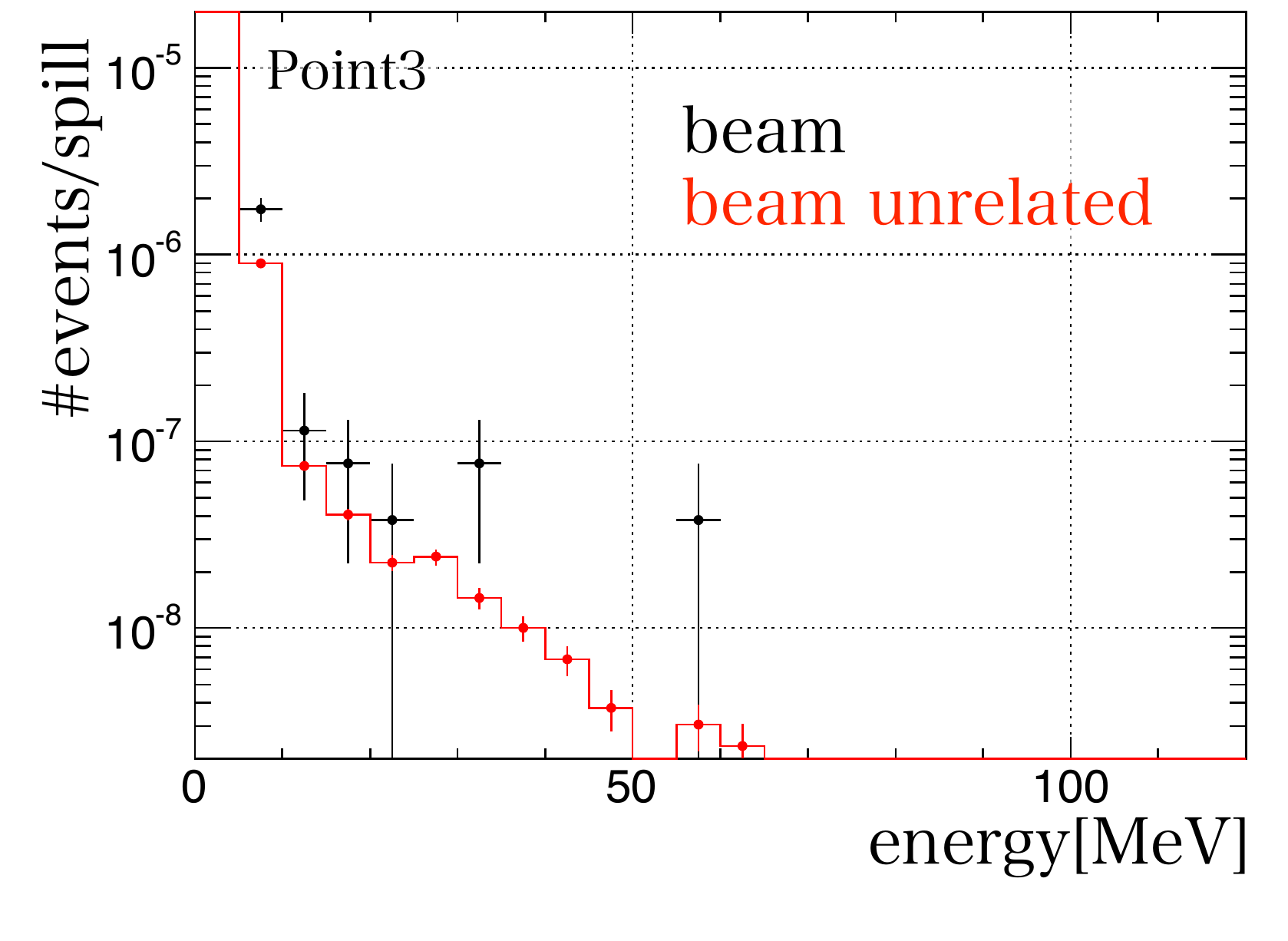}
}
\caption{Energy distributions after applying the on-bunch cut.
The beam unrelated distribution were obtained with the accidental coincidence probability, 3.1\% (``Point~2'') and 0.6$\%$ (``Point~3''), and the energy distribution of beam-off data without applying the on-bunch cut.
} 
\label{500kg_michele_result}
\end{figure}

Figure~\ref{500kg_michele_result} shows the energy distributions after applying the cut to the 500 kg detector data.
The beam unrelated distribution were obtained with the energy distribution of beam-off data without applying the on-bunch cut, and the accidental coincidence probability: hit rate of neutral activities on the bunch timing.
The mean hit rates without veto activities on the bunch timing were 3.1\% (``Point~2'') and 0.6$\%$ (``Point~3'').
The observed event rates during beam-on were $(4.60 \pm 1.53) \times 10^{-7}$/spill (``Point~2'') and $(1.53 \pm 0.76) \times 10^{-7}$/spill (``Point~3''),
while the estimated event rate by beam unrelated activities were $(4.91 \pm 0.28) \times 10^{-7}$/spill (``Point~2'') and $(0.86 \pm 0.04) \times 10^{-7}$/spill (``Point~3''), and both rates between the observed and predicted numbers are consistent.
By considering the efficiency of the on-bunch cut, $\epsilon_{\rm on bunch}=0.9$, the upper limit of the event rate of the beam Michel electron are thus $2.5 \times 10^{-7}$/spill (90\% C.L., ``Point~2'') and $2.1 \times 10^{-7}$/spill (``Point~3'').
Table~\ref{TAB:summary_OBcut} summarizes these numbers. 

\begin{table}[htbp]
\centering
\caption{Summary of the background rates in the prompt region after the on-bunch cut.}
\label{TAB:summary_OBcut}
\begin{tabular}{|c|c|c|}
\hline
& ``Point~2'' ($\times 10^{-7}$/spill) & ``Point~3'' ($\times 10^{-7}$/spill)\\
\hline\hline
Beam-on data &  $4.60 \pm 1.53$ & $1.53 \pm 0.76$ \\ 
Prediction &  $4.91 \pm 0.28$ & $0.86 \pm 0.04$ \\ \hline 
Subtraction & -0.31$\pm$1.56 & 0.67$\pm$0.76\\ \hline 
90$\%$ C.L. upper limit  & $<$ 2.5 & $<$ 2.1 \\ 
\hline
\end{tabular}
\end{table}

\section{Accidental Background}
\indent

Another important background comes from accidental coincidences.
The accidental background rate is estimated using the multiplication
of the single rates of the background in the prompt region and that in the
delayed region. Therefore, an absolute rate of the background
measurements on the prompt and the delayed region with the 500 kg detector are
crucial. Measurements with small size detectors are also important to estimate
the contents (particle identification: PID) of the prompt background.

We first explain the background measurements for the prompt region,
and then for the delayed region.

\subsection{Single Background Rate Measurement for Prompt Region}
\label{SEC:AP}
\indent

As shown in the previous section, the background rates in the prompt
region after charged particle veto are consistent statistically 
between ``beam-on'' and ``beam-off''
data. This means that the beam related background is small enough compared to
the beam unrelated background which is expected to be mainly due to the neutral
 particles, gammas or neutrons induced by the cosmic rays.
Since the response of the E56 detector is quite different for 
gammas and neutrons, it is essential to estimate their influences separately
in the prompt region. 
Thus measurements of the particle identification for the prompt region
was done at Tohoku University using small size detectors at first,
which were used to construct and to check the flux model.   
Finally, the flux model was compared with the 500 kg detector data taken at the
third floor of MLF. All details are described in the following subsections.

\subsubsection{PID and Energy Measurements with Small Size Detectors}
\indent

The calibrations of the small size detectors are described 
elsewhere~\cite{CITE:SR1412}, so only
a summary of the measurements is shown here.  

\begin{figure}
	\centering \includegraphics[width=0.75\textwidth]{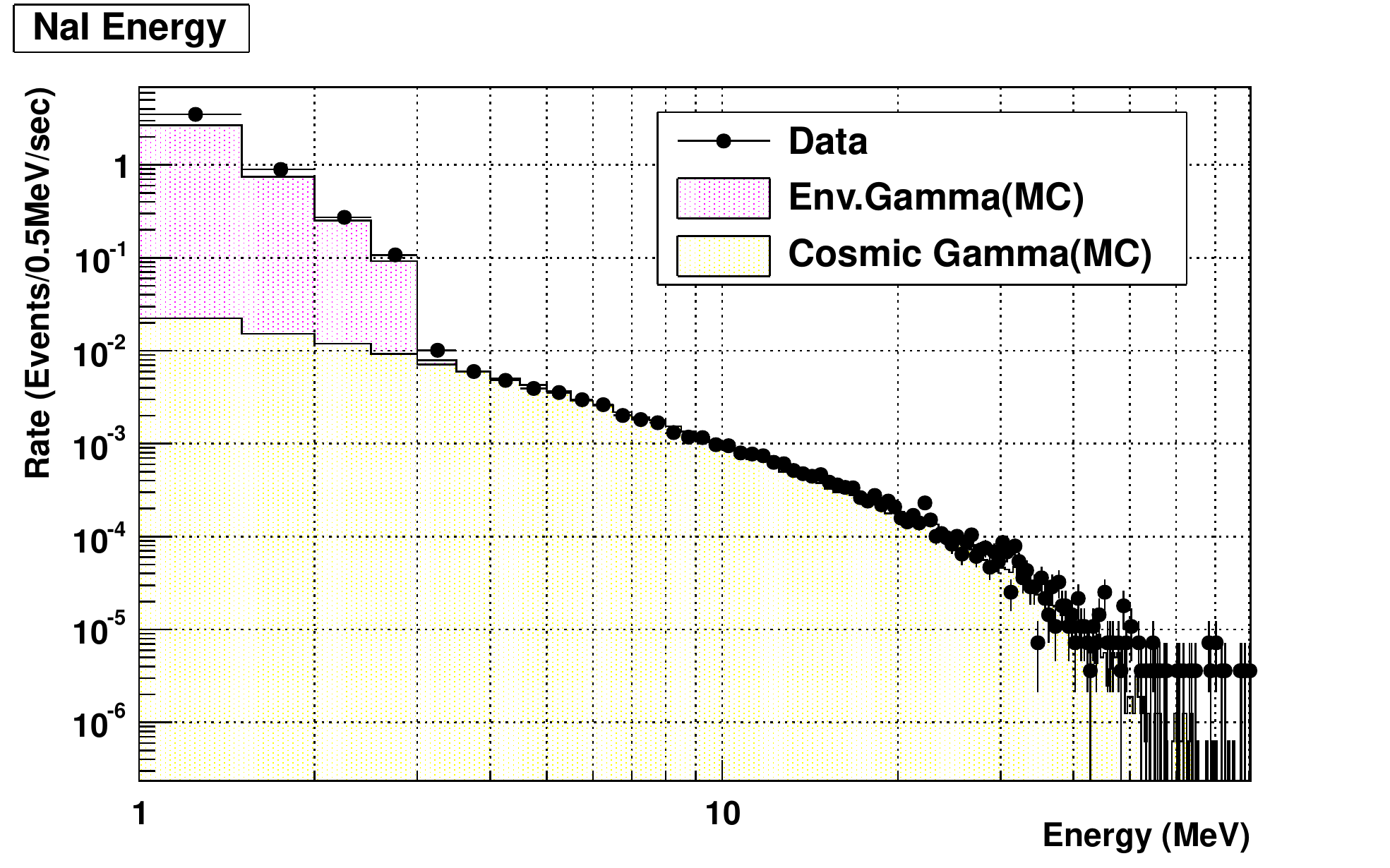}
	\caption{
NaI energy spectrum for the remaining events after applying the veto cut.
The environmental gammas below 3~MeV are generated based on the 
citation~\cite{ENVGGEN}.
Note that the horizontal axis is based on the log-scale, but one bin of the
axis is divided by 0.5 MeV.
}
	\label{fig:NaI_veto2}
\end{figure}

There are no reliable gamma flux measurements or models so far at 
the ground level, which contributes to the prompt event of the accidental 
background. 
The high energy gammas are supposed to be produced mainly 
by the decay of $\pi^0$ which are created by $\mu$-nucleon interactions 
in the walls of the building or air, outside the detectors. 
Therefore we constructed an empirical 
gamma flux model with very simple two exponential functions for energy at 
the surface of the detector at first.
\begin{equation}
\frac{dN}{dE_{n} (MeV)}= \frac{A}{B} \cdot e^{-\frac{E_{n}}{B}} + \frac{C}{D} \cdot e^{-\frac{E_{n}}{D}}
\label{Eq:emp}
\end{equation}
Parameters, $A, B, C, D$ were determined from a Sodium Iodide (NaI) (a cylinder with 2" diameter and 2" height) detector data at Tohoku.
Cosmic ray muons were rejected using 
plastic scintillator anti-counters, which surround the NaI counters, with 
a veto efficiency better than 99$\%$. 
Note that the cosmic $\mu$ veto also eliminated the gammas from 
$\pi^0$s created by $\mu$-nucleon interactions inside the detector. This 
statement can be expanded to any size detectors in general.
Figure~\ref{fig:NaI_veto2} shows the 
results of the measurement, where the energy spectrum in the yellow part can 
be described by two exponential functions with mean energies 
of 3 (``$A$'' in Eq.~\ref{Eq:emp}) and 26 (``$B$'')~MeV (the MC gammas are generated using the two 
exponential functions from the NaI surface assuming an isotropic direction). 
Comparing data and MC spectra, these components rates are measured to be 
150 (``$C$'') for the first exponential function and 25 (``$D$'')~Hz/m$^2$ 
for the second, respectively. 

\begin{figure}[htbp]
\begin{center}
\includegraphics[width=15cm]{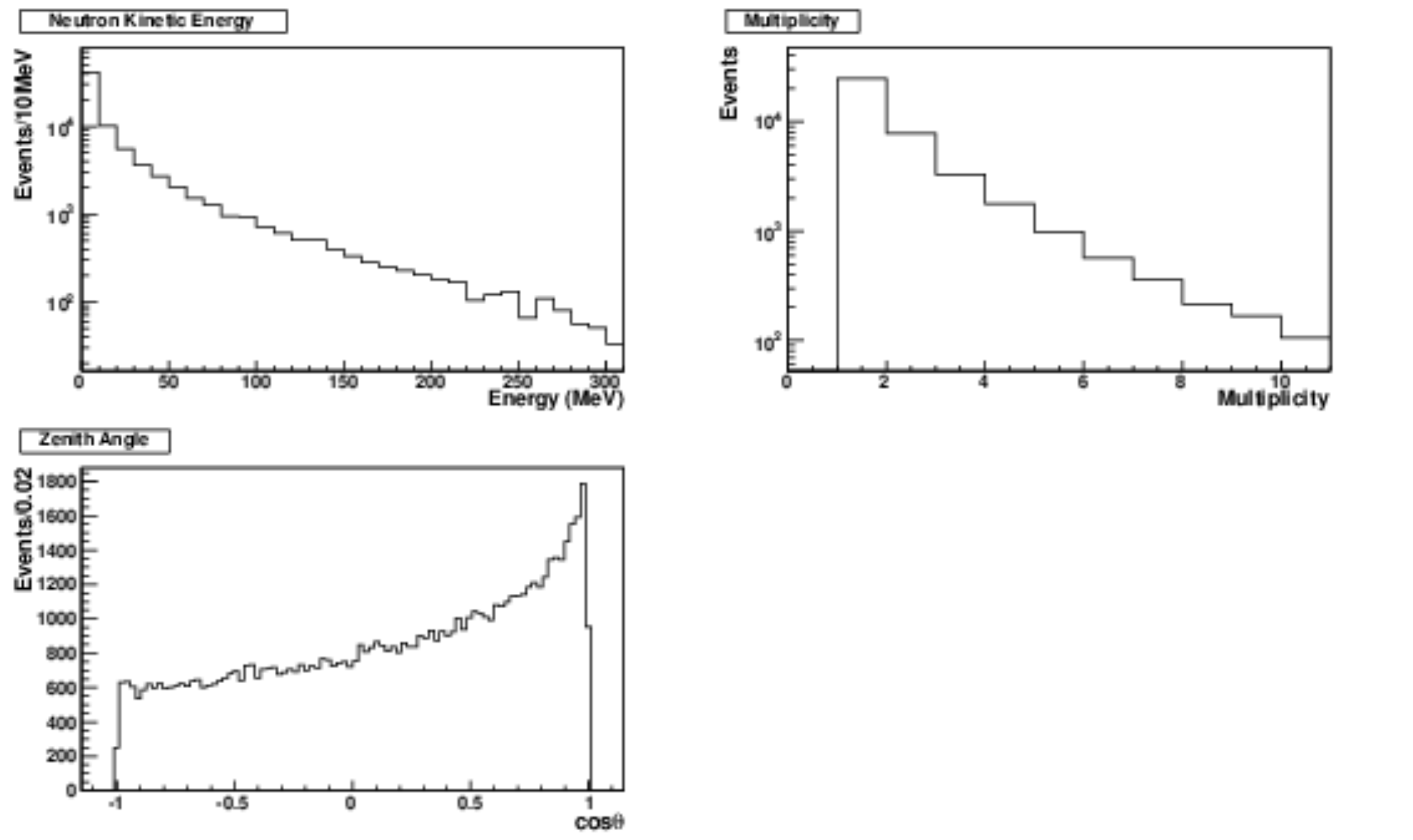}
\end{center}
\caption{\setlength{\baselineskip}{4mm}
	Distributions of the neutron kinetic energy (upper left plot), multiplicity (upper right plot) and zenith angle (lower plot) of the cosmic muon induced fast neutron events generated by the empirical functions in the reference~\cite{COFASTN}.}
\label{COFASTND}
\end{figure}

Another important measurement had been done using a small liquid 
scintillator (NE213~\cite{NE213})
which can separate neutrons from gammas efficiently using a pulse shape of the
scintillation light (PSD: Pulse Shape Discrimination). 
Neutron signals have larger tails than those from gammas, thus 
we can check the flux of fast neutron in addition to the gamma flux which was
measured by NaI already. 
Fast neutrons induced by cosmic rays are measured and modeled by some 
previous works. In this work we employed the model and the parameters 
given by Wang 
{\it et al}~\cite{COFASTN}, who gave  
the empirical functions on the kinetic energy ($E_{n}$), the 
multiplicity ($M$) and the zenith 
angle ($\theta$) of the fast neutrons as a function of the muon 
energy ($E_{\mu}$). 
Figure~\ref{COFASTND} shows those distributions. 

\begin{figure}[htpb!]
	\centering \includegraphics[width=.75\textwidth]{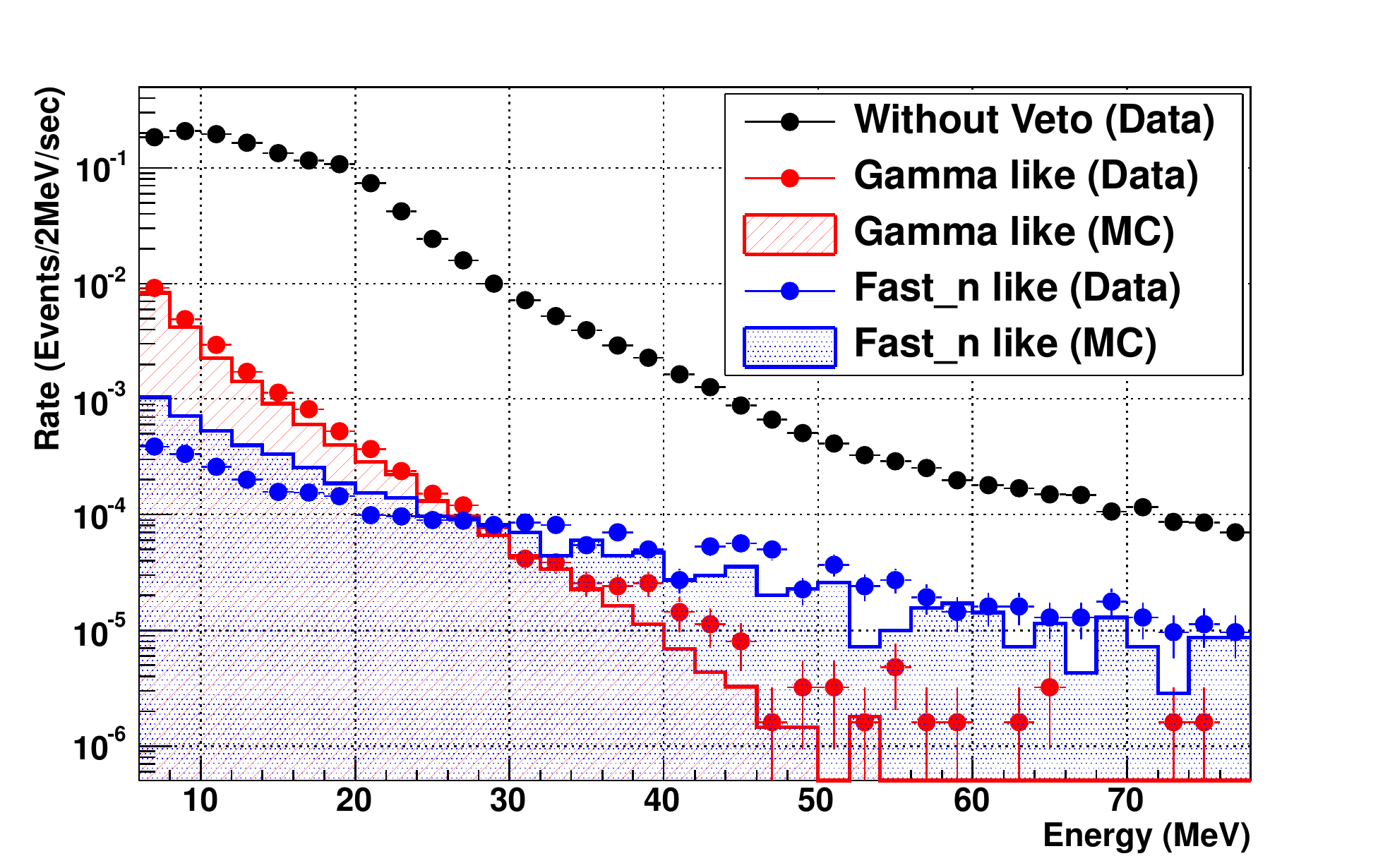}
	\caption{
Reconstructed energy distribution for events with low or no energy deposition in the veto. The data and MC components of neutrons and gammas are also compared. Data events without veto applied are also shown for comparison.}
	\label{fig:BKG_Meigo_Spec2}
\end{figure}

A cylindrical aluminum housing (5" of diameter and 2" height), 
with white painted inner walls, is filled with NE213, closed with a glass plate 
and attached to a 5"  PMT (R1250-03).
The NE213 detector was also surrounded by the plastic scintillators, 
and Fig.~\ref{fig:BKG_Meigo_Spec2} shows the results of the measurement.

With this set-up, the fast neutron flux detected above 20~MeV was of ($1.28 \pm 0.05) \times 10^{-3}$~Hz (statistical uncertainty), while the MC 
based on the reference~\cite{COFASTN} gives a rate of $1.12 \times 10^{-3}$~Hz. 
For gammas, in the same energy range, the measured rate was of ($1.18 \pm 0.04) \times 10^{-3}$~Hz (statistical uncertainty), while the MC based on the NaI measurement described above gives $0.95 \times 10^{-3}$~Hz. MC simulation generates
the gammas at the surfaces of the detectors isotropically.
Therefore, the data and MC above 20~MeV for both gammas and neutrons agrees within 20\% of the uncertainty. 

\subsubsection{Measurement for Prompt Background with 500 kg Detector}
\indent

\begin{figure}[hbtp]
	\begin{center}
		\includegraphics[scale=0.55]{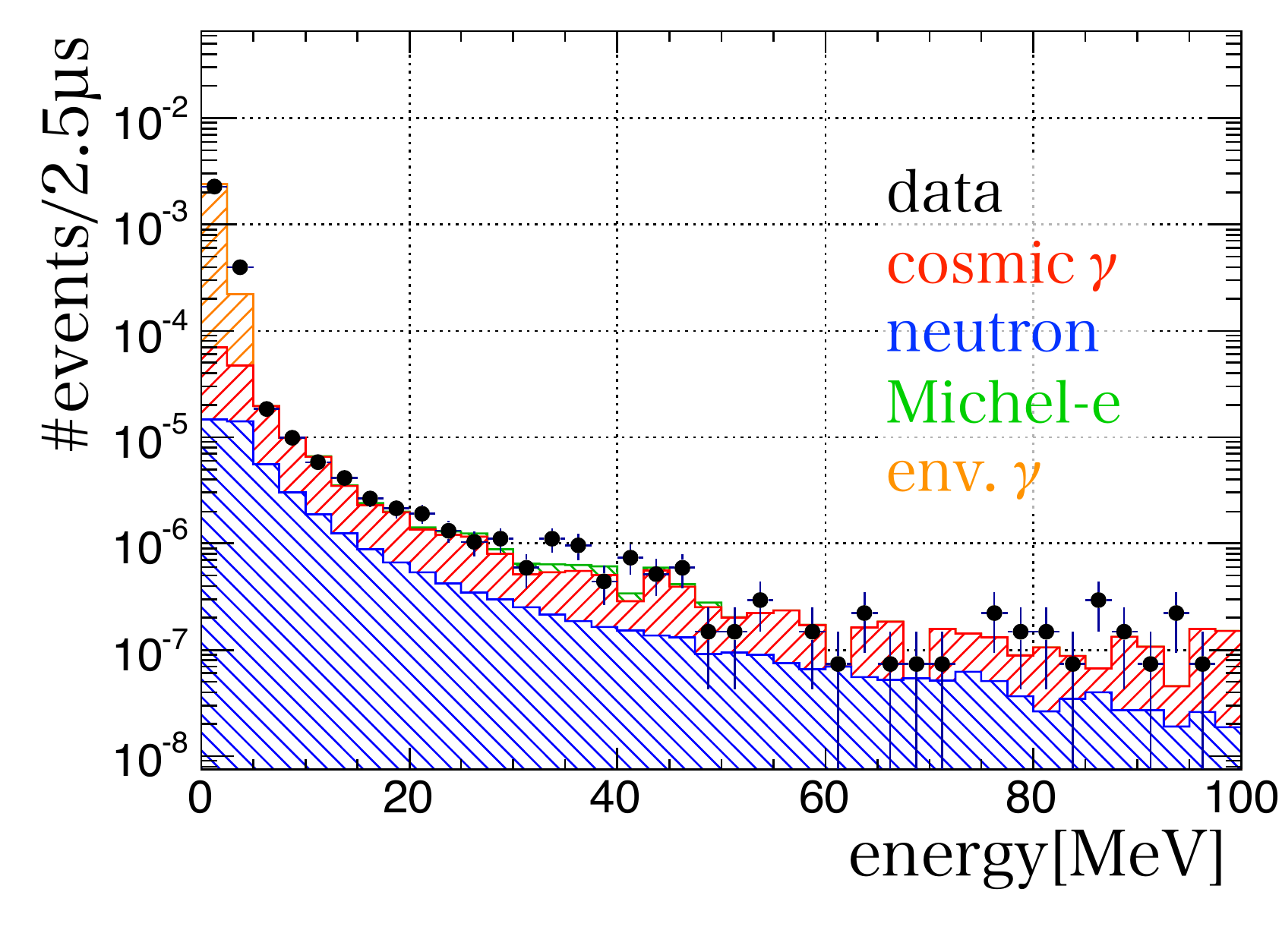}
		\caption{Energy distribution of no-beam data after applying charged particle veto cut and the cosmic Michel electron cut.}
		\label{500kg_prompt_MichelCut}
	\end{center}
\end{figure}

Figure \ref{500kg_prompt_MichelCut} shows the energy distribution of no-beam 
data after applying charged particle veto and the Michel electron cuts. 
Because the activities induced by cosmic rays are dominant in the
prompt region as described above, no beam data is used to estimate the
background for the simplicity.
The MC estimation with cuts is overlaid. 
Fast neutrons are generated based on the model in the
reference~\cite{COFASTN}, and the gammas are generated based on the 
Eq.(\ref{Eq:emp}) at the surface of the 500 kg detector isotropically 
again.

\begin{figure}[hbtp]
	\begin{center}
		\includegraphics[scale=0.55]{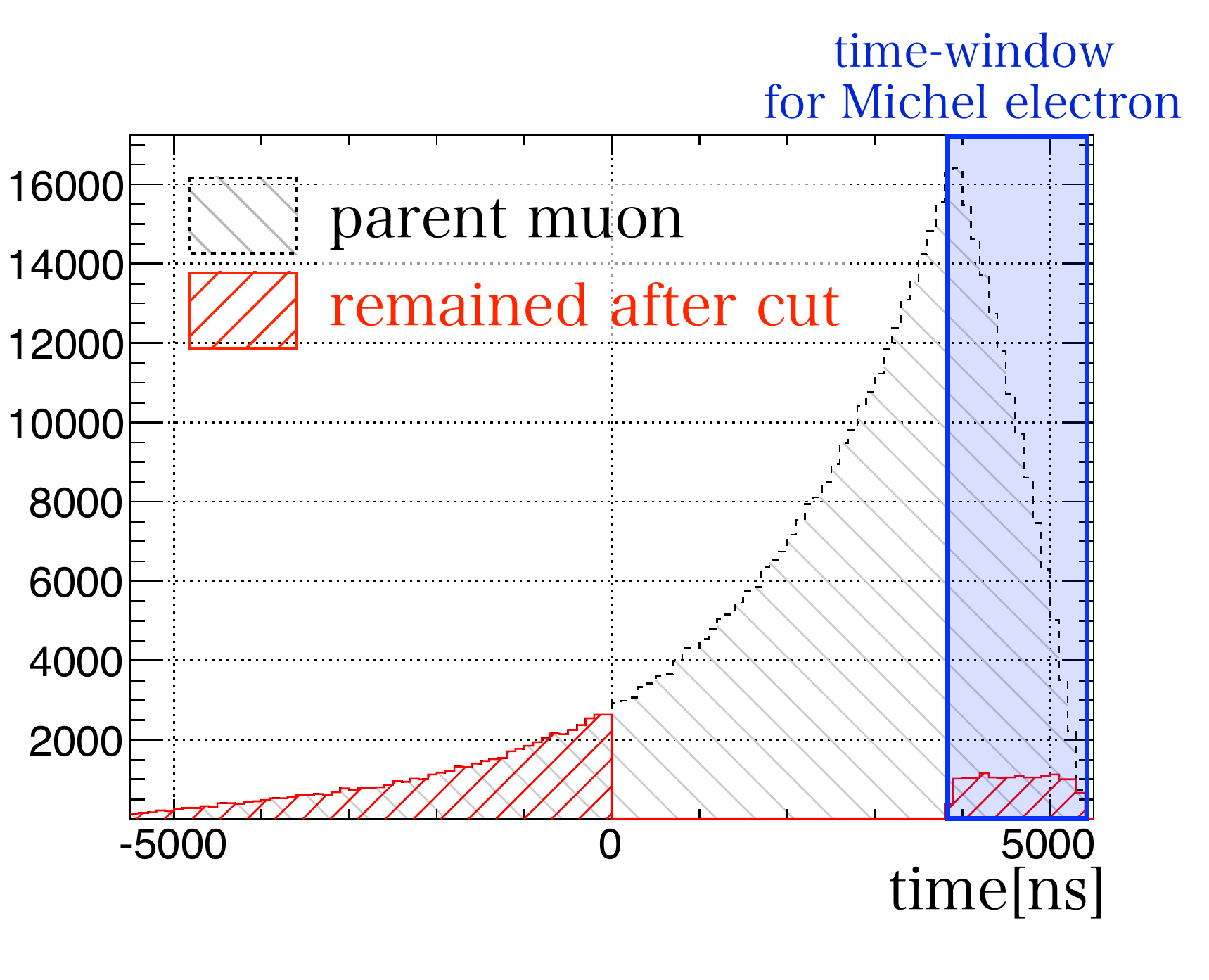}
		\caption{Timing distribution of parent muons, which 
                         generate Michel electrons within a time window, 
                         obtained by a toy Monte Carlo.
		The time window for Michel electrons is $3.9<t[{\rm \mu s}]<5.4$.}
		\label{500kg_prompt_cutEff}
	\end{center}
\end{figure}

The cosmic muons were eliminated by applying the charged particle veto cut because of the high veto efficiency ($>\!99.8$\%) as described in Section~\ref{sec:VetoEff}.
The Michel electrons can be rejected by detecting the parent muons coming earlier than the prompt events by 70~ns. 
Figure \ref{500kg_prompt_cutEff} shows the timing distribution of parent muons, which generate Michel electrons within a time window, $3.9<t[{\rm \mu s}]<5.4$, based on a toy MC. This time window is applied to the real data observation to obtain Fig.~\ref{500kg_prompt_MichelCut}. Some of the events within the time window remained after the cut because the Michel electrons came too close to their parent muons (corresponding to the right red hatched histogram of Fig. \ref{500kg_prompt_cutEff}. )
Events in the negative region also remained because they were out of FADC time window (corresponding to the left red hatched histogram of Fig. \ref{500kg_prompt_cutEff}). 
The rejection power of the Michel electron cut was thus estimated to be 7.
After the cuts, the remaining events are composed of cosmic gammas and neutrons  mainly for the prompt region.

The cosmic induced fast neutrons make the correlated background, however the 25 ton detectors for the E56 will have PID capability, which can reject fast neutrons relative to electrons and gammas by a factor of more than 100~\cite{CITE:P56Proposal}.
Neutrons in the remained events are thus not harmful and only gammas can be a prompt background.
By using gamma flux model described above, 
the number of gammas in the remained events was equivalent 
to $(6.4 \pm 0.5) \times 10^{-6}$ /2.5$\mu$s.

Note that the predicted rate from the measurement at Tohoku 
University was consistent with the rate measured by the 500 kg detector at 
the MLF within 6$\%$. 
As mentioned previously, the goal of the test experiment is to check the 
energy and the rate of the on-site backgrounds. The neutron and gamma flux 
models in this work can be used to discuss the feasibility of the E56
experiment~\cite{CITE:SR1412}.  

\subsection{Single Background Rate Measurement for Delayed Region}
\indent

There are two dominant backgrounds for the delayed region. 
They are neutrons and gammas induced by the beam, and are described in the next subsections.

\subsubsection{Beam Neutrons}
\indent

\begin{figure}
\begin{center}
\includegraphics[width=10cm]{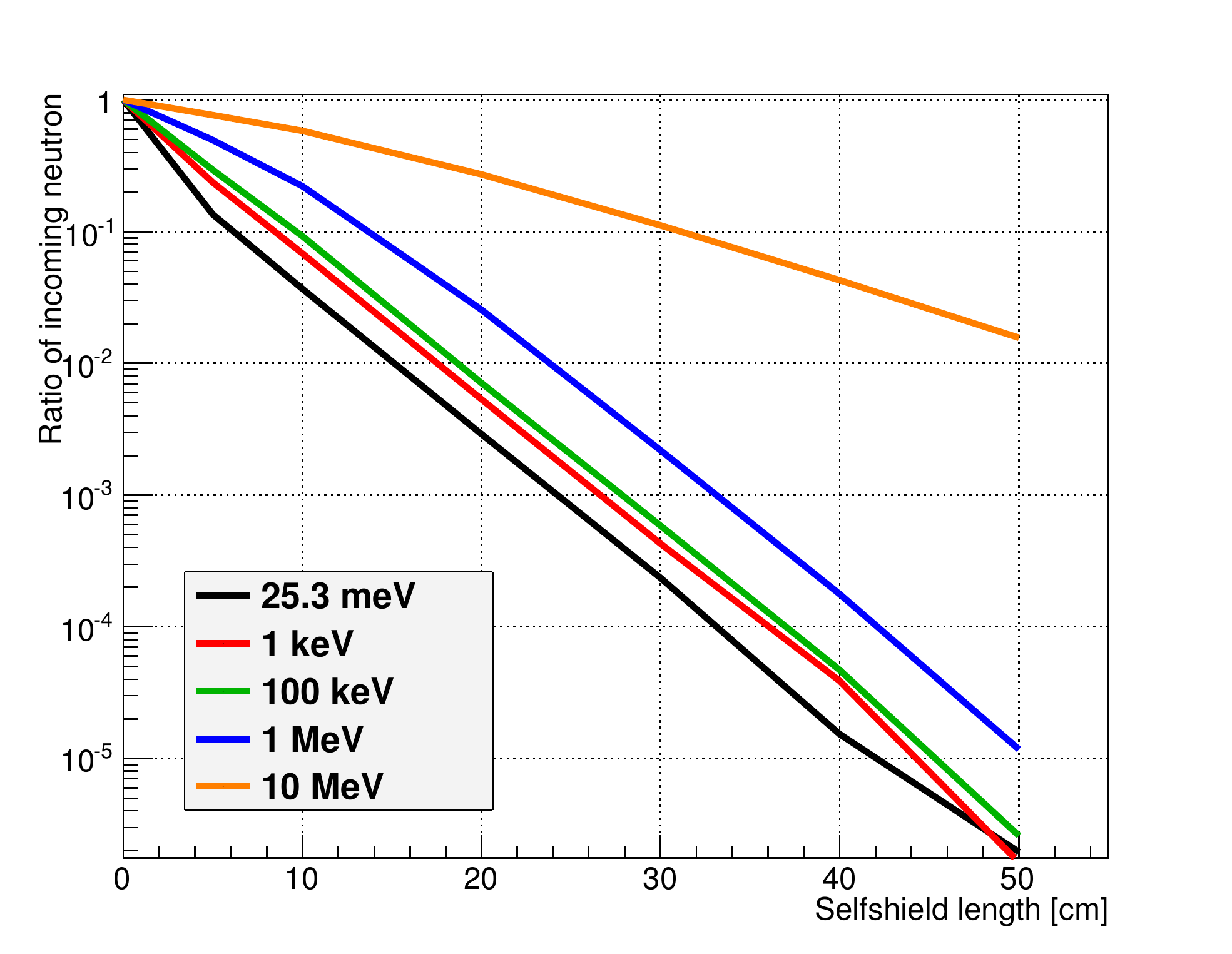}
\caption{Self shield effect of liquid scintillator as function of the shield thickness.}
\label{fig:self_shield_n}
\end{center}
\end{figure}

Thermal neutrons in the third floor of MLF is not harmful for the E56 experiment
since they are stopped at the buffer region of the E56 detector, whose 
thickness of liquid scintillator is 50 cm (Fig.~\ref{FIG:50ton}). But
the middle energy (10 MeV $< E_{kin}$) neutrons on the 
beam bunch can be thermalized inside the detector, and 
can be captured in the liquid scintillator and emits the gammas
inside the fiducial volume. They become backgrounds for the delayed region. 
Figure~\ref{fig:self_shield_n} shows the attenuation rates of thermal neutrons
inside the liquid scintillator generated by Geant4. 
Most of neutrons which have less than 10 MeV are rejected at the buffer region.

\begin{figure}[htpb!]
	\centering \includegraphics[width=.75\textwidth]{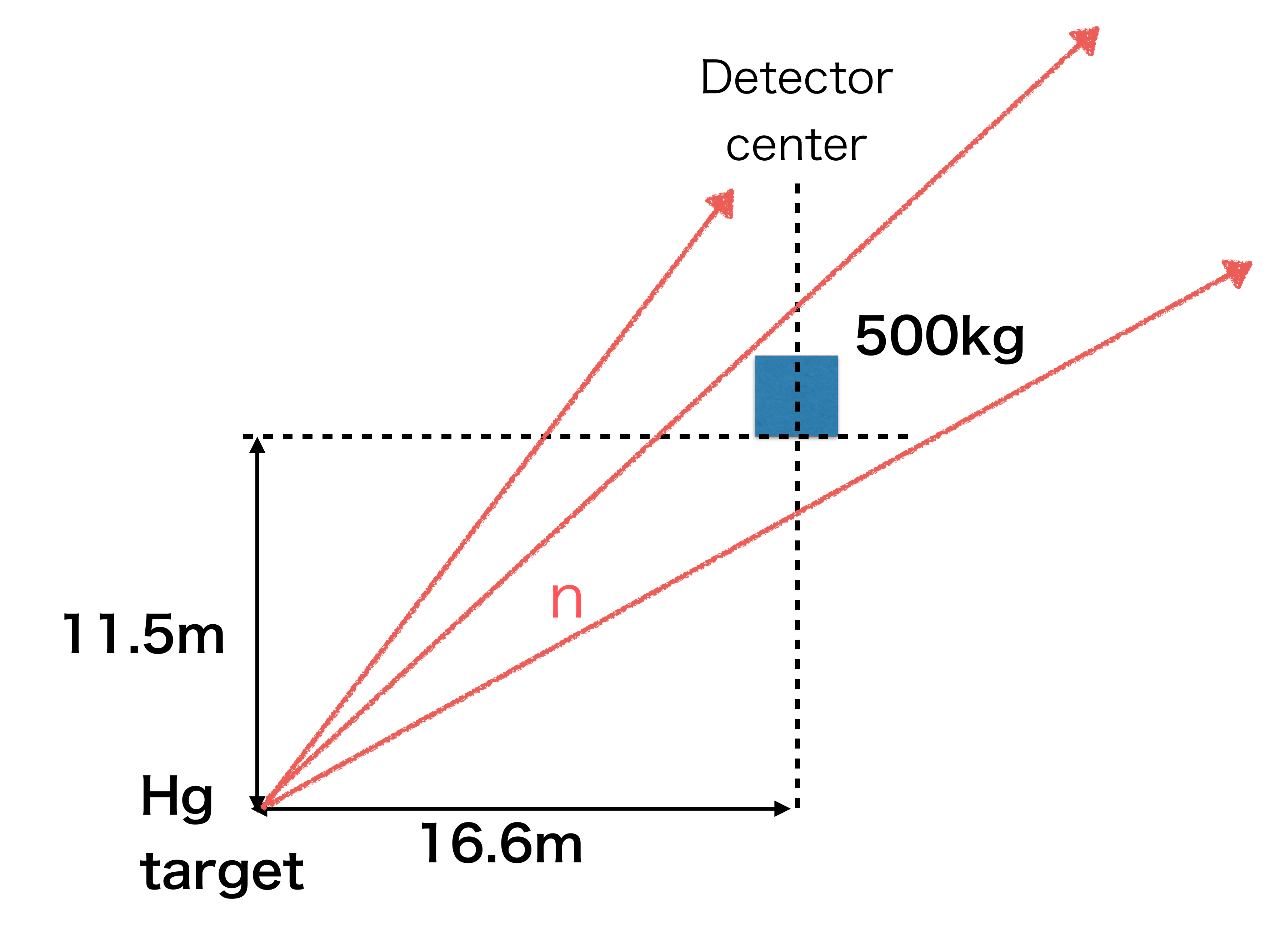}
	\caption{ Geometry of the mercury target and the 500 kg detector.}
	\label{fig:nfromtarget}
\end{figure}

We estimated the neutron flux on the beam bunch timing above 15 MeV using data
assuming the reaction rates of neutrons inside the 500 kg detector based on 
Geant4 since the beam gammas are dominated below 15 MeV as described later. 
Here we used the number of activities and their energy spectrum with the time 
window less than 3 $\mu$s from the beam start time to estimate the flux.
The MC simulation also assumed that such fast neutrons are 
created at the mercury target isotropically and comes to the 
500 kg detector directly.  
Figure~\ref{fig:nfromtarget} shows the schematic view of the assumption. 

This assumption is valid for the neutron backgrounds with only similar solid 
angle positions since the scattered angle, energy and production rate are 
tightly correlated, however the assumption has a good shape for
the ``Point~2'' and ``Point~3'' as described later.

\begin{figure}
\begin{center}
\includegraphics[width=10cm]{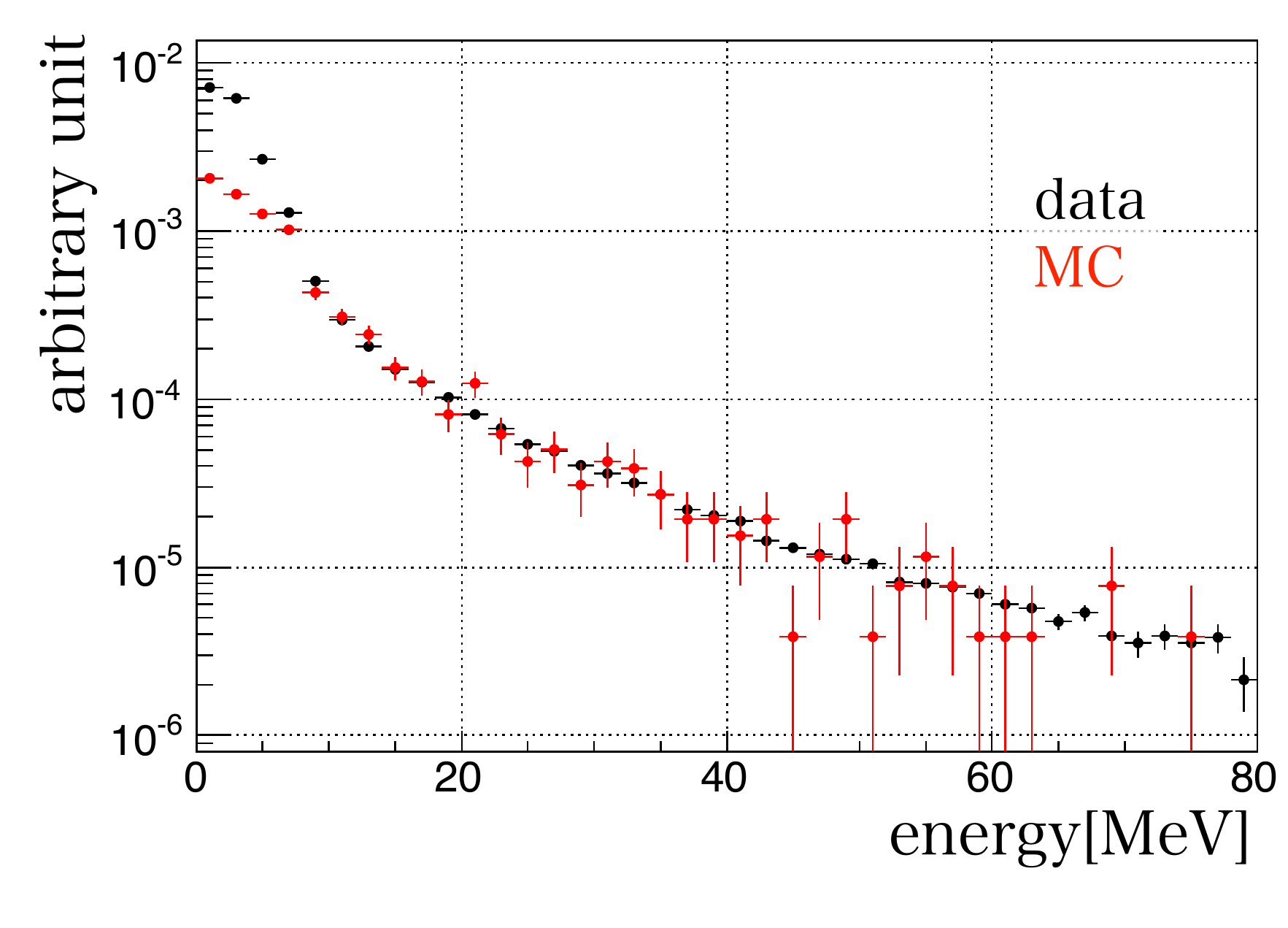}
\caption{
Energy distribution of the 500 kg detector for on-bunch timing. The black cross is the measured energy spectrum and the red one is a MC fitted to the measured one.}
\label{fig:Onbunch_500kg_fit}
\end{center}
\end{figure}

The flux, $\phi(E_{\rm n})$, is denoted as
\begin{equation}
\phi(E_{\rm n}) = \frac{\alpha}{30} \exp(-E_{\rm n}/30), \label{eq:beamNflux}
\end{equation}
where $E_{\rm n}$ is the neutron kinetic energy and $\alpha$ is a number of 
neutrons that are generated at the mercury target in a spill. Note that 
$\alpha$ is just an ``effective'' number of generated neutrons at 
the target
since the 500~kg detector is put on the third floor of MLF, and the neutrons
are highly attenuated by the iron or concrete, which surrounds the target
as shown in Fig.~\ref{FIG:TARGET}.   
$\alpha$ is estimated to be (387$\pm$12)/spill as a fit result at
``Point~2''. 
As shown in Fig.~\ref{fig:Onbunch_500kg_fit}, this flux reproduces well 
the measurement above 15 MeV, which is dominated by activities from neutrons.

The neutron rate at ``Point~2'' with the energy above 15 MeV 
is (5.4$\pm$0.02)$\times$10$^{-3}$/spill/0.3MW/500kg, 
while that at ``Point~3'' is \\
 (2.0$\pm$0.01)$\times$10$^{-3}$/spill/0.3MW/500kg. 
Although the neutron flux depends on 
the scattering angle from the mercury target,
these are well explained by the 1/$r^{2}$ law, where $r$ is the 
detector baseline, within 5$\%$.

\subsubsection{Beam Gammas}
\indent

Beam gammas are produced by capture reactions of the thermalized beam neutrons
at the materials of MLF building such as concrete walls or floors.
They are injected into the E56 detector, therefore they are 
crucial background in the delayed region.

\begin{figure}
\begin{center}
\includegraphics[width=10cm]{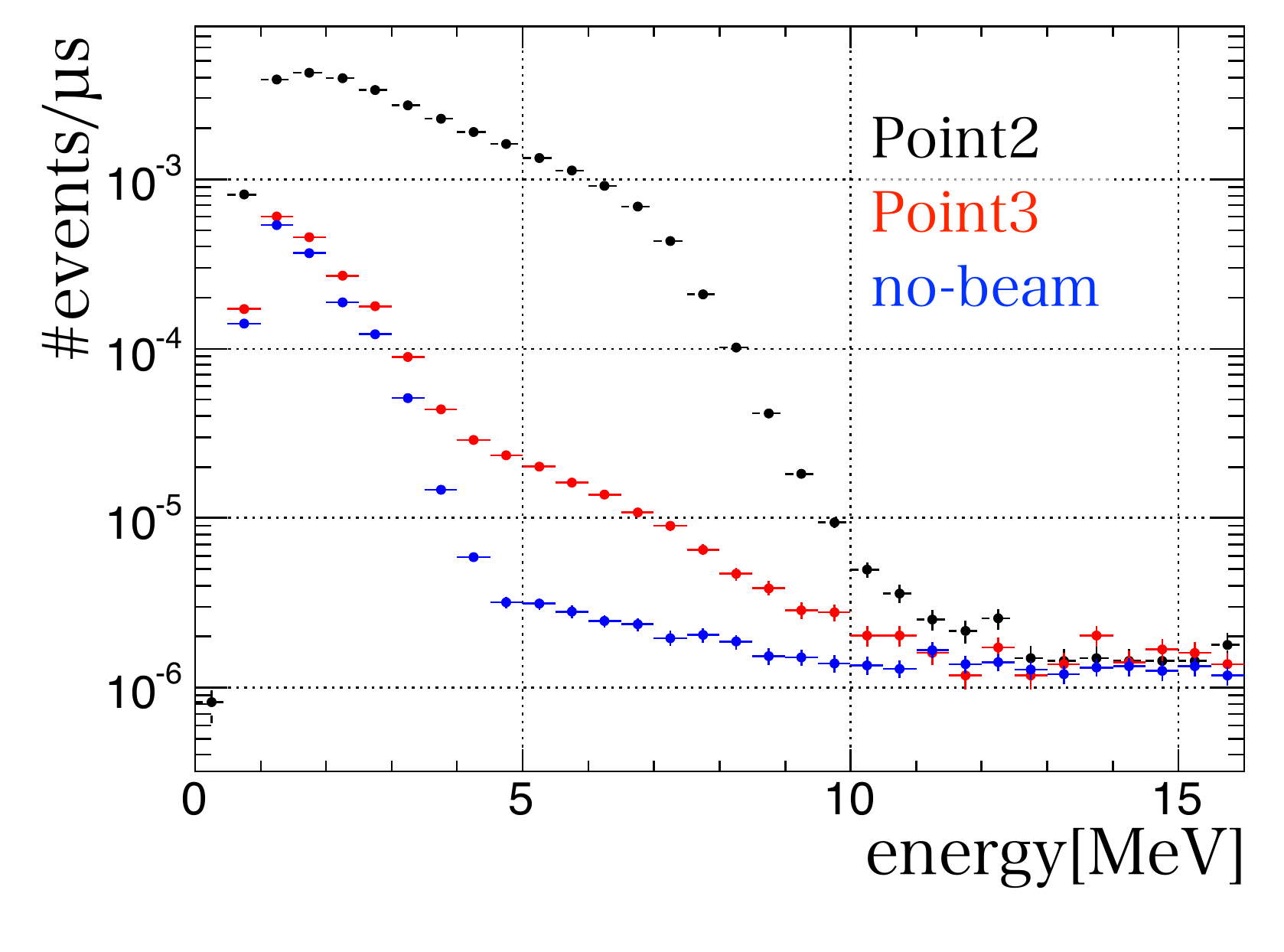}
\caption{
Energy distribution of the backgrounds for delayed region measured with the 500 kg detector. Black: energy spectrum at ``Point~2'', red: that at ``Point~3'', blue: no beam data.}
\label{fig:Delayed_gamma_measurement}
\end{center}
\end{figure}

As shown in Fig.~\ref{500kg_TvsE}(a) and (b), 
the hit time distributions of the low energy activities were almost flat.
Therefore even the rate measurements during short time span such as 
only 1~$\mu$s is useful to estimate the background rates for 
the IBD delayed region.  
Figure~\ref{fig:Delayed_gamma_measurement} shows the energy spectra below 
16 MeV measured by the 500 kg detector. 
$4.35<t[{\rm \mu s}]<5.35$ from the rising edge of the first beam bunch 
shown in Fig.~\ref{500kg_TvsE}(a) and (b) were used to 
estimate background in the IBD delayed region. 
In the energy region of 7 $< E [{\rm MeV}] <$ 12, 
the background rate with the 500 kg detector are
$(8.26\pm0.07)\times 10^{-4}$/$\mu$s/0.3MW at ``Point~2'', and
$(3.65\pm0.12)\times 10^{-5}$/$\mu$s/0.3MW at ``Point~3''.

\begin{figure}
\begin{center}
\includegraphics[width=12cm]{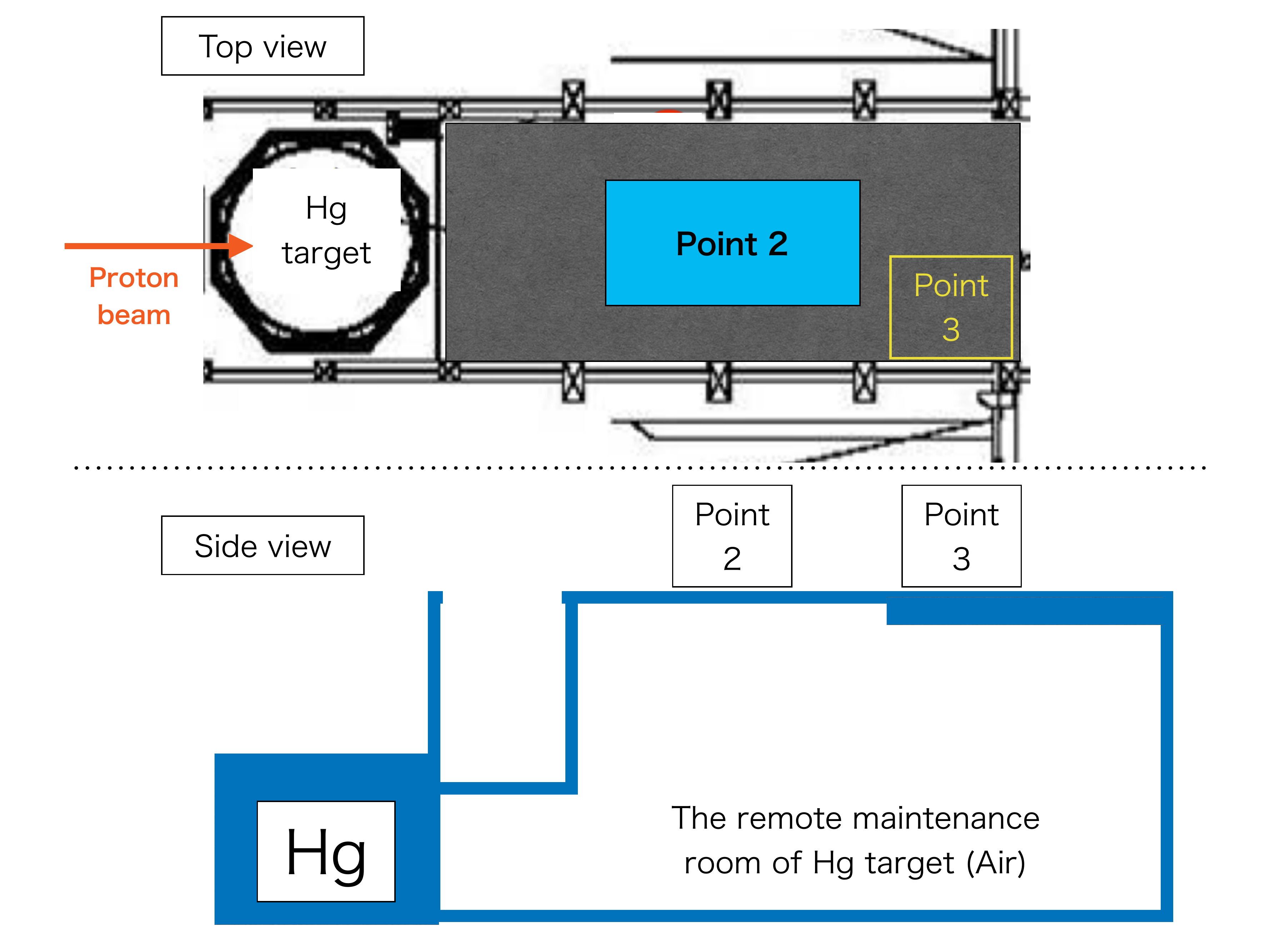}
\caption{
The top view of the third floor of MLF around the measurement points. 
Bottom is the side view. 
The floor of ``Point~2'' is made of concrete which has 120 cm thickness and is thinner than that of ``Point~3'' by 30 cm.}
\label{fig:Delayed_premeasurement}
\end{center}
\end{figure}

As shown in the bottom plot of Fig.~\ref{fig:Delayed_premeasurement}, 
there is a room for the remote maintenance of the mercury target 
under ``Point~2'' and ``Point~3''. 
The floor of ``Point~2'' is made of concrete which has 120 cm thickness and is thinner than that of ``Point~3'' by 30 cm.
This structure of the MLF building can make the 
difference of the background rates between ``Point~2'' and ``Point~3''. 

\section{Summary}
\indent

The J-PARC MLF 2013BU1013 test experiment was carried out to measure the backgrounds 
and to examine the feasibility of the J-PARC E56 experiment.
The results show that:
\begin{itemize} 
\item Beam fast neutrons:\\ 
The beam fast neutrons which create charged pions 
inside a 500 kg plastic scintillators, are not observed during 2 weeks. 
The 90$\%$ C.L. upper limit is set to be 2.5$\times$10$^{-7}$/2.9$\mu$s/0.3MW at
 ``Point~2'', and 2.1$\times$10$^{-7}$/2.9$\mu$s/0.3MW at ``Point~3''. 
\item Accidental:
\begin{enumerate} 
\item The gammas or neutrons induced by cosmic rays are observed for the prompt 
region. The neutron rate is the same as expected in the 
reference~\cite{COFASTN}. Gammas are newly recognized by 
this experiment. 
The number of gammas
is equivalent to $(6.4 \pm 0.5) \times 10^{-6}$ /2.5$\mu$s in the 500 kg 
detector in the prompt region.
\item Neutrons and gammas induced by the beam are observed for the delayed 
region by the 500~kg detector. The flux of neutrons are estimated
by the data, and the shape of the flux is consistent with the exponential 
with a slope constant of 30 MeV. The effective neutron production rate is 
(387$\pm$12)/spill at the target. ``Point~2'' and ``Point~3'' have same neutron 
flux within 5$\%$, considering the baseline effect.
The rates of gammas are 
$(8.26\pm0.07)\times 10^{-4}$/$\mu$s/0.3MW (``Point~2'') and 
$(3.65\pm0.12)\times 10^{-5}$/$\mu$s/0.3MW (``Point~3'') in the energy region of
7 $< E [{\rm MeV}] <$ 12.  
\end{enumerate}
\end{itemize}  

These information have been translated to the background rates at 
the E56 detectors using the MC simulation with evaluation of 
the self-shielding 
effects, and the results show that the backgrounds at the ``Point~2'' 
provide no specific problems to perform the J-PARC E56
experiment~\cite{CITE:SR1412}.

\section{Acknowledgements}
\indent

We warmly thank the MLF people, especially, Prof. Masatoshi Arai, 
who is the MLF Division leader, and 
the neutron source group, muon group and user facility group for the 
various kinds of support.
The LEPS2 experiment has lent us the good quality scintillators, electronics, cables, PMTs, and we deeply appreciate it.
We borrow cables, electronics, scintillators from J-PARC Hadron 
group, University of Kyoto, JAEA, KEK online group, Belle II 
experiment and T2K experiment and we would like to express appreciation 
for their kindness.
We also thank Bill Louis (LANL), Minfang Yeh (BNL) and Joshua Spitz (MIT)
for the useful discussion.

Finally, we thank the support from J-PARC and KEK.  


\end{document}